\documentclass[aps,twocolumn,prd,nofootinbib,showpacs,unsortedaddress]{revtex4}

\usepackage{amsmath}
\usepackage{amssymb}
\usepackage{graphicx}

\newcommand{\sss}[1]{{\scriptscriptstyle{#1}}}
\newcommand{\vect}[1]{\mathbf{#1}}
\newcommand{\heaviside}[1]{\mathrm{H}\!\left( #1 \right)}
\newcommand{\CAMB}{\texttt{CAMB}}
\newcommand{\GHz}{\textrm{GHz}}
\newcommand{\Gpc}{\textrm{Gpc}}
\newcommand{\zero}{{\sss{0}}}
\newcommand{\uCMB}{\mathrm{\sss{CMB}}}
\newcommand{\ulss}{\mathrm{lss}}
\newcommand{\uinf}{\mathrm{inf}}
\newcommand{\ures}{\mathrm{res}}
\newcommand{\ufov}{\mathrm{fov}}
\newcommand{\um}{\mathrm{m}}
\newcommand{\ub}{\mathrm{b}}
\newcommand{\uc}{\mathrm{c}}
\newcommand{\us}{\mathrm{s}}
\newcommand{\un}{\mathrm{n}}
\newcommand{\ud}{\mathrm{d}}
\newcommand{\ui}{\mathrm{i}}
\newcommand{\ue}{\mathrm{e}}
\newcommand{\uh}{\mathrm{h}}
\newcommand{\uD}{\mathrm{D}}
\newcommand{\umin}{{\min}}
\newcommand{\umax}{{\max}}
\newcommand{\ucoh}{\mathrm{coh}}
\newcommand{\ustg}{\mathrm{stg}}
\newcommand{\uall}{\mathrm{all}}
\newcommand{\Tcmb}{T_\uCMB}
\newcommand{\LCDM}{\Lambda\mathrm{CDM}}
\newcommand{\ksilk}{k_\uD}
\newcommand{\kperp}{l}
\newcommand{\vbaryon}{v_{\ub}}
\newcommand{\unitn}{\vect{\hat{n}}}
\newcommand{\aexp}{a}
\newcommand{\boxmpc}{L_\mathrm{sim}}
\newcommand{\corrini}{\ell_\uc}
\newcommand{\horizonini}{d_{\uh_\zero}}
\newcommand{\newton}{G}
\newcommand{\tension}{U}      
\newcommand{\GU}{\newton \tension}
\newcommand{\OmegaM}{\Omega_\um}
\newcommand{\angx}{\alpha}
\newcommand{\angy}{\beta}
\newcommand{\ang}{\theta}
\newcommand{\angres}{\ang_\ures}
\newcommand{\angfov}{\ang_\ufov}
\newcommand{\nablanum}{\nabla_{\!\un}}
\newcommand{\powspec}{P}
\newcommand{\uwav}{u}

\begin{document}

\title{Small-angle CMB temperature anisotropies induced by cosmic
  strings}


\author{Aur\'elien A. Fraisse}
\email{fraisse@astro.princeton.edu}
\affiliation{Princeton University Observatory, Peyton Hall, Princeton,
  New Jersey 08544, USA}

\author{Christophe Ringeval}
\email{ringeval@fyma.ucl.ac.be}
\affiliation{Theoretical and Mathematical Physics Group, Center for
  Particle Physics and Phenomenology, Louvain University, 2 Chemin du
  Cyclotron, 1348 Louvain-la-Neuve, Belgium}

\author{David N. Spergel}
\email{dns@astro.princeton.edu}
\affiliation{Princeton University Observatory, Peyton Hall, Princeton,
  New Jersey 08544, USA\\
  Princeton Center for Theoretical Physics, Princeton, New Jersey
  08544, USA}

\author{Fran\c{c}ois R. Bouchet}
\email{bouchet@iap.fr}
\affiliation{Institut d'Astrophysique de Paris, UMR CNRS 7095,
  Universit\'e Pierre et Marie Curie, 98bis boulevard Arago, 75014
  Paris, France}


\date{August 25, 2008}


\begin{abstract}
We use Nambu-Goto numerical simulations to compute the cosmic
microwave background~(CMB) temperature anisotropies induced at
arcminute angular scales by a network of cosmic strings in a
Friedmann-Lema\^{i}tre-Robertson-Walker (FLRW) expanding universe.  We
generate $84$ statistically independent maps on a $7.2^\circ$ field of
view, which we use to derive basic statistical estimators such as the
one-point distribution and two-point correlation functions.  At high
multipoles, the mean angular power spectrum of string-induced CMB
temperature anisotropies can be described by a power law slowly
decaying as $\ell^{-p}$, with $p=0.889$ ($+0.001,-0.090$) (including
only systematic errors).  Such a behavior suggests that a
nonvanishing string contribution to the overall CMB anisotropies may
become the dominant source of fluctuations at small angular scales.
We therefore discuss how well the temperature gradient magnitude
operator can trace strings in the context of a typical arcminute
diffraction-limited experiment.  Including both the thermal and
nonlinear kinetic Sunyaev-Zel'dovich effects, the Ostriker-Vishniac
effect, and the currently favored adiabatic primary anisotropies, we
find that, on such a map, strings should be ``eye visible,'' with
at least of order ten distinctive string features observable on a
$7.2^\circ$ gradient map, for tensions $\tension$ down to $\GU \simeq
2\times 10^{-7}$ (in Planck units).  This suggests that, with upcoming
experiments such as the Atacama Cosmology Telescope~(ACT), optimal
non-Gaussian, string-devoted statistical estimators applied to
small-angle CMB temperature or gradient maps may put stringent
constraints on a possible cosmic string contribution to the
CMB~anisotropies.
\end{abstract}


\maketitle


\section{Introduction}
\label{sec:intro}

The idea that all elementary-particle forces could simply be different
manifestations of a single underlying force gained much strength after
the successful elaboration of a unified renormalizable theory of
electromagnetic and weak forces~\cite{Weinberg:1980}.  If the quest
for the ultimate grand unified group of symmetries $G$ has so far been
unsuccessful, the mere idea of unification leads to strong theoretical
predictions.  In particular, if the particle physics interactions were
unified at high energy, the expansion of the Universe should have
triggered spontaneous breakdown of their
symmetries~\cite{Kirzhnits:1972}.  Kibble showed that the induced
phase transitions may form stable topological
defects~\cite{Kibble:1976}.  Cosmic strings are the linelike version
of such primordial vacuum remnants.  Their energy per unit length
$\tension$ is directly linked to the energy scale of the phase
transition during which they were formed.  Moreover, once formed,
cosmic strings are stable and should still be present nowadays.
Although an inflationary era occurring after a string-forming phase
transition would dilute the defects enough to render them
unobservable, defects formation at the end of inflation is a generic
feature of particle physics motivated scenarios~\cite{Jeannerot:2003}.
Motivated by the coincidence that 
strings formed at the grand unified theory (GUT) energy scale would
induce density fluctuations with amplitude close to the observed
amplitude of galaxy fluctuations, they were once considered as serious
candidates to explain structure formation in our
Universe~\cite{Durrer:2002}.  However, the COBE data, associated to
large-scale structure observations, already suggested that
adiabatic fluctuations seeded galaxy formation~\cite{Albrecht:1997},
and the most recent high precision measurements of the cosmic
microwave background (CMB) anisotropies leave little doubt that this
conclusion is right~\cite{Spergel:2007}.  Quite recently, cosmic
strings have also attracted the interest of the fundamental string
theory community: the embedding of inflation in string theory may
indeed produce another class of stringlike cosmological objects,
dubbed ``cosmic superstrings,'' that could be fundamental or Dirichlet
strings associated with extradimensional branes.  Although of
intrinsic Planckian-like energy density, their effective
four-dimensional mass would be redshifted in presence of warped extra
\nobreak{dimensions}~\cite{Davis:2005}.

These theoretical considerations motivate the study of string-induced
gravitational effects, such as lensing or gravitational wave
emission~\cite{Kaiser:1984, Gott:1985, Vilenkin:1981, Damour:2001,
  Siemens:2006, Bernardeau:2001, Oguri:2005, Seljak:2006}.  Using CMB
and large-scale structure data, an upper limit to a potential cosmic
defects contribution to the primordial inhomogeneity can be
derived~\cite{Wyman:2005, Fraisse:2007}.  The corresponding upper
bound on the dimensionless energy scale $\GU$ (in Planck units) of
local strings currently ranges from $2\times 10^{-7}$ to $10^{-6}$ and
is a weak function of the model used to describe the string network.
With the notable exception of Abelian string networks, whose induced
CMB power spectrum has been recently obtained on the currently
observable angular scales~\cite{Bevis:2007}, many of the CMB analyses
performed so far use analytical or semianalytical defect models which
may or may not accurately mimic the cosmological evolution of a string
network.  These analytical approximations are often introduced to
circumvent the difficulties associated with the highly nonlinear
evolution of cosmic strings in an expanding universe.  The theoretical
understanding of cosmic string evolution in a
Friedmann-Lema\^{i}tre-Robertson-Walker (FLRW) universe is still an
active field of research which has led to the development of numerical
simulations incorporating all the defect dynamics~\cite{Bennett:1989,
  Albrecht:1989, Bennett:1990, Allen:1990, Vincent:1998, Moore:2002,
  Vanchurin:2005, Martins:2006, Ringeval:2007}.  A drawback is that
they have a limited dynamic range in redshift making it impossible to
directly compute the evolution of a network of strings from their
formation to the present time.

These numerical limitations have motivated the search for simple
signatures in the cosmological observables, such as straight
temperature steps in the CMB temperature maps~\cite{Jeong:2007}.  With
the advent of new arcminute CMB experiments, such as the Arcminute
Microkelvin Imager~(AMI)~\cite{Ami:2006}, the South Pole Telescope
(SPT)~\cite{Ruhl:2004} or the Atacama Cosmology Telescope
(ACT)~\cite{Kosowsky:2006}, looking for strings directly into these
maps may be promising.  However, the efficiency of such searches would
be greatly enhanced by the theoretical knowledge of the exact string
patterns that may be imprinted in the microwave sky.

The goal of this paper is to generate realistic simulated CMB
temperature maps that include the effect of cosmic strings on
arcminute angular scales.  For this purpose, we use high-resolution
numerical simulations of Nambu-Goto strings based on an improved
version of the Bennett and Bouchet
code~\cite{Bennett:1990,Ringeval:2007}.  As already mentioned,
numerical simulations cannot probe a very wide range of redshifts.
But as far as small angular scales are concerned, we can use the
approach introduced by Bouchet~\cite{Bouchet:1988} and compute the
strings' evolution starting at the last scattering surface.  Our method
allowed the generation of 84 statistically independent maps, which we
used to extract some basic statistical properties associated with the
string patterns.  Special attention has been paid to quantifying the
systematic errors induced by the numerical initial conditions.  We
find that applying the gradient magnitude operator to our simulated
CMB temperature maps leads to the reconstruction of the string shapes
on the past light cone, which suggests that it may be of some help for
direct searches.  We then add Gaussian perturbations of inflationary
origin and the Sunyaev-Zel'dovich~(SZ) effects to our simulated maps.
Convolving the resulting maps with a diffraction-limited beam based on
the specifications of the ACT experiment, we find that the gradient
magnitude operator exhibits the strings' signature down to $GU=2\times
10^{-7}$.  While we defer the use of more sophisticated statistical
estimators to a future paper, the clear non-Gaussian patterns showing
up in the maps suggest that these estimators will significantly
improve the current constraints on $GU$.

The paper is organized as follows.  In Sec.~\ref{sec:simmaps}, we
present our numerical simulations of cosmic string networks and the
method used to simulate pure string-induced CMB temperature maps.
Sec.~\ref{sec:stats} is devoted to a basic statistical analysis
conducted on $84$ simulated maps.  We give the mean probability
distribution function of the temperature fluctuations induced by
strings and the corresponding mean angular power spectrum, with an
estimation of the various associated errors.  Some properties of the
gradient magnitude of the string-induced temperature anisotropies are
also discussed.  In Sec.~\ref{sec:obs}, we examine the observability
of strings for an ACT-like experiment when the primary anisotropies,
SZ effects, and the effect of beam smoothing are taken into account.
The suitability of the gradient maps as strings tracer is reexamined
in this context.  We present our conclusions~in~Sec.~\ref{sec:end}.


\section{Simulated maps}
\label{sec:simmaps}

A variety of methods have been used to compute the CMB anisotropies
induced by cosmic strings.  Many approaches rely on the use either of
Green functions~\cite{Hindmarsh:1995} or of unequal time correlators
(UTC)~\cite{Pen:1997}, and all of them require the knowledge of the
defect stress tensor evolution during the cosmological expansion.
Because of the intrinsic nonlinear evolution of defect networks, this
is usually achieved through numerical simulations in FLRW
space-time~\cite{Bennett:1989, Albrecht:1989, Allen:1990,
  Vincent:1998, Moore:2002}, although simple analytical defect
models~\cite{Durrer:1996, Zhou:1996, Albrecht:1999, Pogosian:1999}
have widely been used to circumvent the numerical limitations
mentioned in Sec.~\ref{sec:intro}.

Computing the corresponding CMB angular power spectra requires only
knowledge of the defect stress tensor two-point correlation functions.
Since topological defects are active and incoherent sources of gravity
perturbations, their two-point functions are generically nonvanishing
at unequal times~\cite{Magueijo:1996}.  The UTC method relies on the
scaling properties associated with an evolving cosmological string
network to make this calculation easier.  The network scaling
properties indeed strongly restrict the UTC's functional form.  They
are then fit to numerical simulations and extrapolated to cosmological
scales.  This method has been successfully applied to derive the CMB
and matter power spectra for a variety of topological
defects~\cite{Durrer:1999, Contaldi:1999, Wu:2002, Bevis:2004} and has
recently been used with Abelian strings in Ref.~\cite{Bevis:2007}. 

Since we are interested in generating realistic maps, we need to go
beyond computing the power spectrum.  The two-point functions do not
encode the non-Gaussian features induced by networks of defects in the
CMB.  We therefore use numerical simulations to capture the nonlinear
effects associated with the defect evolution, including
non-Gaussianity.  This approach has already been applied to constrain
the energy breaking scale associated with various cosmic defects and
to simulate full-sky CMB maps for Nambu-Goto strings~\cite{Pen:1994,
  Allen:1997, Landriau:2003, Landriau:2004}.  But due to the rather
small expansion factor numerically achievable, such maps can only
include stringy effects up to a finite redshift, typically
$z \simeq 10^2$.  The CMB anisotropies computed in this way are
therefore only accurate on large angular scales.  We avoid this
limitation by stacking maps from different redshifts, an approach
outlined in Ref.~\cite{Bouchet:1988} and applied in
Ref.~\cite{Bennett:1992}.  While simulations with the observer outside
of the numerical box are not well suited for a full-sky map
reconstruction, they are quite useful for the small angular scales
considered in this paper.

Denoting $\Theta_\ell$ the $\ell$th multipole moment of the
temperature perturbation to photon distribution, the Boltzmann
hierarchy in Fourier space can be recast into~\cite{Dodelson:2003,
  Peter:2005}
\begin{equation}
\label{eq:losight}
\begin{aligned}
\Theta_\ell = & \int_0^{\eta_\zero} g(\eta)\,\ue^{-k^2/\ksilk^2}
  \left(\bar{\Theta}_\zero + \bar{\Phi} \right)
  j_\ell(k \Delta \eta)\, \ud \eta \\[1mm]
            & + \int_0^{\eta_\zero} \eta\, g(\eta)\,
  \ue^{-k^2/\ksilk^2}\, \mathrm{i}\, \bar{v}_\mathrm{b}\,
  j_\ell'(k\Delta\eta)\, \ud \eta \\[1mm]
            & + \int_0^{\eta_\zero} \ue^{-\tau}
  \left(\dfrac{\ud \Phi}{\ud \eta} + \dfrac{\ud \Psi}{\ud \eta}\right)
  j_\ell(k\Delta\eta)\, \ud \eta,
\end{aligned}
\end{equation}
where $\eta_\zero$ stands for the present conformal time, $\Psi$ and
$\Phi$ are the Bardeen potentials, $\Delta \eta = \eta_\zero - \eta$,
and $\vbaryon$ refers to the baryon velocity~\cite{Bardeen:1980}.  A
bar has been used when the Silk damping term has been explicitly
extracted from the perturbation variables~\cite{Silk:1968}.  The
visibility function $g$ is related to the optical depth $\tau$ through 
\begin{equation}
g(\eta) \equiv - \dfrac{\ud \tau}{\ud \eta}\, \ue^{-\tau},
\end{equation}
and is strongly peaked around $\eta_\ulss$, the time of last
scattering.  As a result, the contribution of the first two integrals
in Eq.~(\ref{eq:losight}), namely, the Sachs-Wolfe and Doppler terms,
is only significant for $\eta \simeq \eta_\ulss$~\cite{Sachs:1967}.
Moreover, Silk damping exponentially washes these terms out for
$k > \ksilk$, i.e., typically for $\ell \gtrsim 2000$.  In presence of
a cosmological string network, the Bardeen potentials $\Psi$ and 
$\Phi$ in Eq.~(\ref{eq:losight}) are sourced by all the usual
cosmological fluids, as well as by the strings.  Following
Ref.~\cite{Durrer:2002}, if $\Psi_\ustg$ and  $\Phi_\ustg$ refer to
the perturbations that would be sourced by the strings alone, one can
define $\Psi_\um \equiv \Psi - \Psi_\ustg$ and
$\Phi_\um \equiv \Phi - \Phi_\ustg$.  As only a minor string
contribution appears to be compatible with the current CMB
data~\cite{Wyman:2005, Fraisse:2007}, $\Psi_\ustg \ll \Psi$, 
$\Phi_\ustg \ll \Phi$, and one may consider the stringy effects as a
first order correction to the standard adiabatic cosmological
perturbations.  Although, in the presence of strings, the Bardeen
potentials $\Psi_\um$ and $\Phi_\um$ are coupled to $\Psi$ and $\Phi$
through the perturbed Einstein equations, at leading order,
\begin{equation}
\label{eq:stgpert}
\Psi_\um \simeq \Psi_\ucoh \quad \mathrm{and} \quad
\Phi_\um \simeq \Phi_\ucoh,
\end{equation}
where the index ``coh'' refers to the purely coherent Bardeen
potentials of inflationary origin obtained without strings.
Therefore, dropping the Sachs-Wolfe and Doppler terms in
Eq.~(\ref{eq:losight}), one gets
\begin{equation}
\begin{aligned}
\label{eqn:thetalkgg1}
\Theta_\ell \underset{k\gg1}{\simeq} &
  \int_0^{\eta_\zero} \ue^{-\tau}
  \big(\dot{\Phi}_\ucoh + \dot{\Psi}_\ucoh \big)
  j_\ell(k\Delta\eta)\, \ud \eta \\[1mm]
            & + \int_0^{\eta_\zero} \ue^{-\tau}
  \big(\dot{\Phi}_\ustg + \dot{\Psi}_\ustg \big)
  j_\ell(k\Delta \eta)\, \ud \eta,
\end{aligned}
\end{equation}
where a dot denotes a derivative with respect to $\eta$.  The first
term represents the integrated Sachs-Wolfe effect for the standard
cosmological fluids and would vanish in a purely matter-dominated
universe.  However, due to the recent domination of the cosmological
constant and the existence of radiation residuals at the surface of
last scattering, this term contributes significantly to the large
angular scales, but is unimportant at small
ones~\cite{Dodelson:2003}.  Moreover, as the Universe is optically
thick before last scattering, Eq.~(\ref{eqn:thetalkgg1}) becomes
\begin{equation}
\label{eq:stgisw}
\Theta \simeq \int_{\eta_\ulss}^{\eta_\zero} \ue^{-\tau}
  \big(\dot{\Phi}_\ustg + \dot{\Psi}_\ustg\big)
  \ue^{-\mathrm{i}\, \vect{k}\cdot \vect{x}_\gamma}\, \ud \eta,
\end{equation}
where $\vect{x}_\gamma(\eta) \equiv \unitn\, \Delta\eta$ is the photon
path along the line of sight.  At small angular scales, one may
therefore expect the strings' signature in the CMB temperature
fluctuations to be dominated by their integrated Sachs-Wolfe~(ISW)
effect from the last scattering surface.  In the following, we use
cosmic string numerical simulations and the so-called small-angle
approximation to simulate CMB temperature maps according to
Eq.~(\ref{eq:stgisw}).


\vspace{0.10cm}
\subsection{Small-angle approximation}

For Nambu-Goto strings, $\Phi_\ustg$ and $\Psi_\ustg$ are solution of
the perturbed Einstein equations sourced by the Nambu-Goto stress
tensor.  In the temporal gauge ($X^0=\eta$), with
$\alpha \equiv U/\sqrt{-g}$, it reads 
\begin{equation}
\label{eq:ngstress}
T^{\mu\nu} = \alpha \int \ud \sigma 
  \left(\epsilon\, \dot{X}^\mu \dot{X}^\nu - 
        \dfrac{1}{\epsilon}\, X^{\prime \mu} X^{\prime \nu}\right)
  \delta^3 \left(\vect{x} - \vect{X}\right),
\end{equation}
where $\epsilon^2 \equiv \vect{X}'^2/(1-\dot{\vect{X}}^2)$, $U$ is the
string energy per unit length entering the definition of the
Nambu-Goto action, and a prime denotes a derivative with respect to
the string world sheet spacelike coordinate
$\sigma$~\cite{Vilenkin:2000}.  In the small-angle approximation,
which is well suited for angles typically smaller than the Hubble
angular size at the epoch of interest, Hindmarsh, Stebbins and
Veeraraghavan showed that, in the case of Nambu-Goto strings,
Eq.~(\ref{eq:stgisw}) can be simplified to~\cite{Hindmarsh:1994,
  Stebbins:1995}
\begin{equation}
\label{eq:stgsa}
\Theta \simeq 
  \dfrac{8\pi\mathrm{i}\, \newton \tension}{\vect{\kperp}^2}
  \int_{\vect{X}\,\cap\,\vect{x}_\gamma}
  \left(\vect{u} \cdot \vect{X}\right)
  \ue^{-\mathrm{i}\, \vect{\kperp} \cdot {\vect{X}}}\,
  \ue^{-\tau}\epsilon\, \ud \sigma.
\end{equation}
The wave vector $\vect{l}$ denotes the transverse component
of~$\vect{k}$ with respect to the line of sight $\unitn$, whereas, in
the temporal gauge, $\vect{u}$ encodes the string stress tensor
distortions of the photon temperature and reads 
\begin{equation}
\label{eq:udef}
\vect{u} = \dot{\vect{X}} - 
  \dfrac{(\unitn \cdot \vect{X}') \cdot \vect{X}'}{1 +
  \unitn \cdot \dot{\vect{X}}}.
\end{equation}
As can be seen in Eq.~(\ref{eq:stgsa}), only the strings that
intercept the photon path can imprint their signature in the CMB
temperature fluctuations.  As a result, the knowledge of $\vect{u}$,
and therefore of the string trajectories $\vect{X}$, is only required
on our past light cone.  In the context of string numerical
simulations, the trajectories of all the strings are computed for all
times.  Therefore, to compute $\vect{u}$, one only needs to determine
what parts of the string network intercept our past light cone at a
given time.  Equation~(\ref{eq:stgsa}) was also at the basis of the
earliest string maps presented in Ref.~\cite{Bouchet:1988}.  However,
our source terms in Eq.~(\ref{eq:udef}) are slightly different than
the ones used in Refs.~\cite{Bouchet:1988, Stebbins:1988} and include
an additional \nobreak{longitudinal} component with respect to the string
trajectory~\cite{Hindmarsh:1994}.  As~discussed in
Ref.~\cite{Stebbins:1995}, this component encodes a logarithmic
correction to the temperature fluctuations induced by the string
curvature and may be significant for wiggly strings.


\vspace{0.10cm}
\subsection{Cosmic string evolution}
\label{sect:csevol}

Since Eq.~(\ref{eq:stgisw}) involves an integration from the last
scattering surface to the present time, we are left with simulating
the cosmological evolution of a network of cosmic strings from
$z\simeq 1100$ to $z\simeq 0$. 

Our numerical simulations in FLRW space-time are based on an improved
version of the Bennett and Bouchet Nambu-Goto cosmic string
code~\cite{Bennett:1990, Ringeval:2007}.  The runs are performed in a
comoving box with periodic boundary conditions and whose volume has
been scaled to unity.  The scale factor is initially normalized to
unity, while the horizon size $\horizonini$ is a free parameter
controlling the initial string energy within a horizon volume.  During
the computation, the comoving horizon size grows and the evolution is
stopped before it fills the whole unit volume to avoid spurious
effects coming from the boundary conditions.  We use the
Vachaspati-Vilenkin initial conditions for which the long strings
path is essentially a random walk of correlation length $\corrini$,
together with a random transverse velocity component of root mean
squared amplitude $0.1$~\cite{Vachaspati:1984}.  The initial horizon
size and velocity amplitude are chosen to minimize the relaxation time
of the Vachaspati-Vilenkin string network toward its stable
cosmological configuration.  As mentioned before, the finite numerical
box limits the range of scale factors during which the string
evolution can be computed at once.  In our case, to model the network
evolution since the epoch of last scattering, we would need to follow
it over a period during which the Universe expands by a factor of
order $10^3$, which is not achievable, even with the current
supercomputing facilities.  Moreover, applying the results of the
small-angle approximation would turn out to be difficult in such a
case.  Indeed, if $z_\ui$ denotes the redshift at which we start the
numerical computation, the numerical box corresponds to a real
comoving size 
\begin{equation}
\label{eq:boxmpc}
\boxmpc = \dfrac{2/\horizonini}{\aexp_\zero H_\zero \sqrt{\OmegaM}
                                \sqrt{1 + z_\ui}},
\end{equation}
in the matter-dominated era and for a flat universe.  The density
parameter $\OmegaM$ refers to the matter content of the Universe
today, whereas $\horizonini$ is expressed in units of the numerical
box size.  For the largest simulations to date, $z_\ui = 1089$ and
$\horizonini \simeq 5\times 10^{-2}$ are typical values of these two
parameters (see, e.g., Ref.~\cite{Ringeval:2007}).  In a flat $\LCDM$
universe, using fiducial values for the density parameters compatible
with the three-year Wilkinson Microwave Anisotropy Probe (WMAP)
data~\cite{Spergel:2007}, this gives $\boxmpc \simeq 9\,\Gpc$ and the
numerical box then subtends an angular size of
$\angfov \simeq 27^\circ$.  The largest structures in such a map would
therefore be out of the small-angle limit we are interested in.  As a
result, we adopt the approach introduced in Ref.~\cite{Bouchet:1988},
which relies on two smaller runs.  The first one starts at the last
scattering surface and ends at a redshift fixed by the maximum
expansion factor achievable in the numerical box.  For the simulations
we performed, $\horizonini \simeq 0.185$, which corresponds to
$\boxmpc\simeq 1.7\,\Gpc$ and $\angfov \simeq 7.2^\circ$.  Such a run
ends after a 30-fold increase in expansion factor, corresponding to a
redshift $z \simeq 36$.  We then propagate the photons perturbed by
the first run into a second numerical simulation of the same size but
starting at $z_\ui\simeq 36$.  For another 30-fold increase in
expansion factor, this run ends at $z\simeq 0.3$.  As can be seen in
Eq.~(\ref{eq:boxmpc}), the second simulation represents a much larger
real volume than the first one and therefore subtends a greater angle
in the sky.  As a result, only the subpart of the second run that
matches the angle subtended by the first simulation is actually used.
As we will see later on, the CMB temperature maps are weakly sensitive
to the string network at low redshifts, simply because there are
almost no strings intercepting our past light cone in a recent past,
which makes this technique perfectly acceptable. 

In practice, each of these numerical simulations is started before the
redshifts we mentioned in order to give the cosmic string network
enough time to relax toward its stable cosmological configuration.  As
soon as this scaling regime is established, the string characteristic
lengths evolve in a self-similar way with respect to the horizon size,
allowing the numerical results to be rescaled to cosmological
distances.  Therefore, one has to make sure that the structures
(strings and loops) we are interested in have indeed reached their
scaling behavior during the numerical runs.  This is not an issue for
the so-called infinite (or long) strings, defined as strings larger
than the horizon size, because they rapidly reach the scaling regime.
Although it has been shown in Ref.~\cite{Ringeval:2007} that the
cosmic string loop distribution scales as well, the relaxation time
for the loops to reach such a self-similar evolution with respect to
the horizon size appears to be larger for smaller loops.  As a result,
and this is inherent to all cosmic string numerical simulations, the
smaller length scales in a numerical string network keep some memory
of the initial network configuration until they reach their stable
cosmological evolution (see also Refs.~\cite{Martins:2006,
  Vanchurin:2006, Olum:2007}).  Note that even if this memory effect
is physical, one does not expect a physical string network at the last
scattering surface to still exhibit structures coming from its initial
configuration at the GUT energy scale.  The change in scale factor
between the GUT redshift and the last scattering surface is indeed so
huge that all the observable length scales should be in scaling at
decoupling.  Unfortunately, as already mentioned, numerical
simulations cannot probe the evolution of a cosmic string network up
to the GUT epoch.  To circumvent this issue, we switch on the photon
propagation inside the runs only after making sure all the large
structures (infinite strings and loops) are in their scaling regime.
This can be checked, for instance, by monitoring the evolution of the
energy density distributions.  In the following, we start the photons' 
propagation when all loops larger than a third of the horizon size are
in scaling.  This cutoff is then dynamically pushed toward smaller
values to include all the loops entering the scaling regime at later
times.  The cutoff time dependence can be deduced from the loop
distribution relaxation times derived in the simulations performed in
Ref.~\cite{Ringeval:2007}.  We discuss the residual systematic errors
associated with the presence of nonscaling structures in
Sec.~\ref{sec:tempmaps}.  Finally, let us point out that our
Nambu-Goto simulations are not smoothing out the so-called
``small-scale structure'' along the strings~\cite{Bennett:1990}.  As a
result, the CMB maps we generate depend on only one parameter, namely,
the string energy per unit length $\tension$.


\vspace{0.10cm}
\subsection{Temperature fluctuation maps}
\label{sec:tempmaps}

\begin{figure*}
\begin{center}
\includegraphics[width=7.cm]{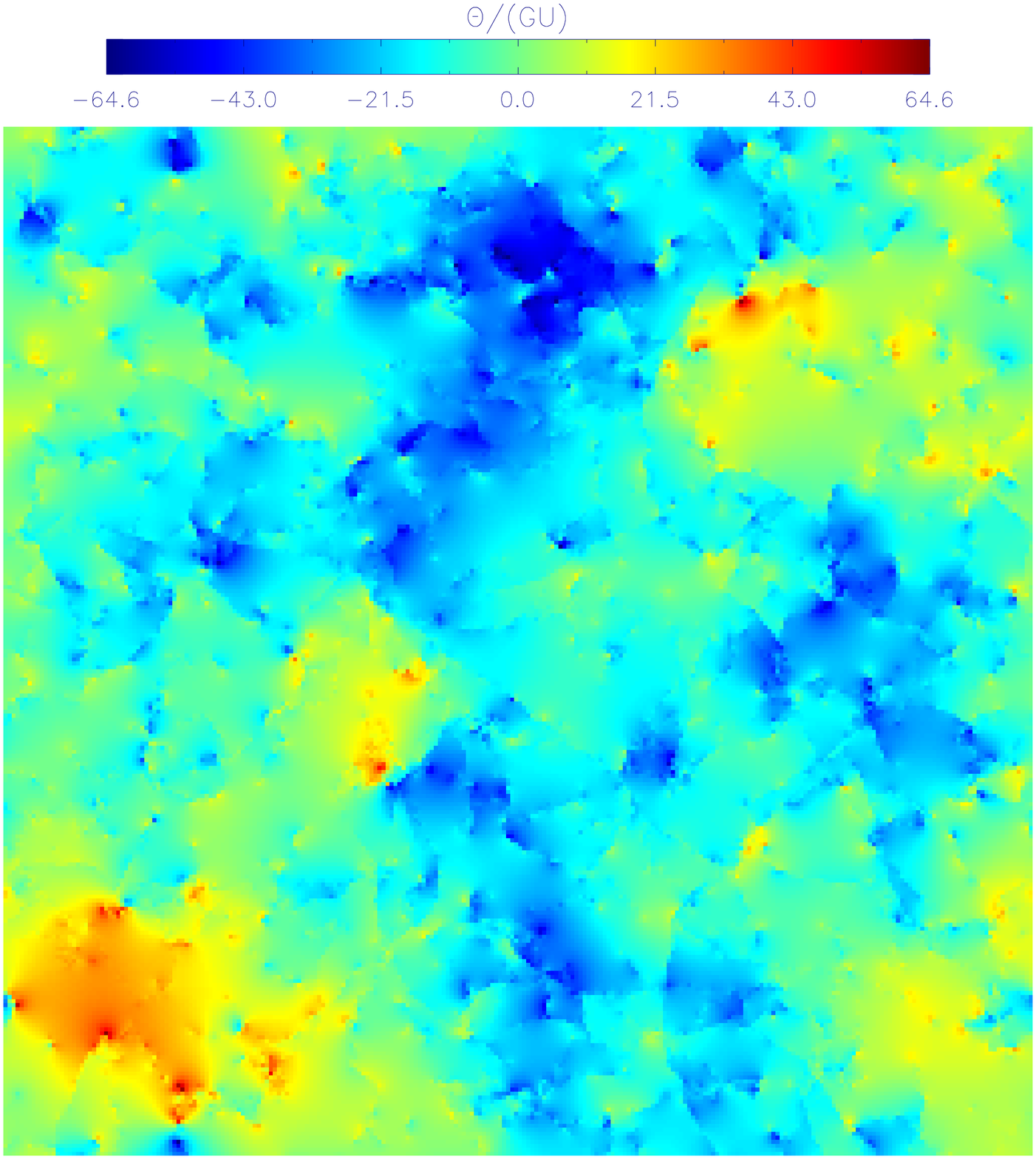}
\includegraphics[width=7.cm]{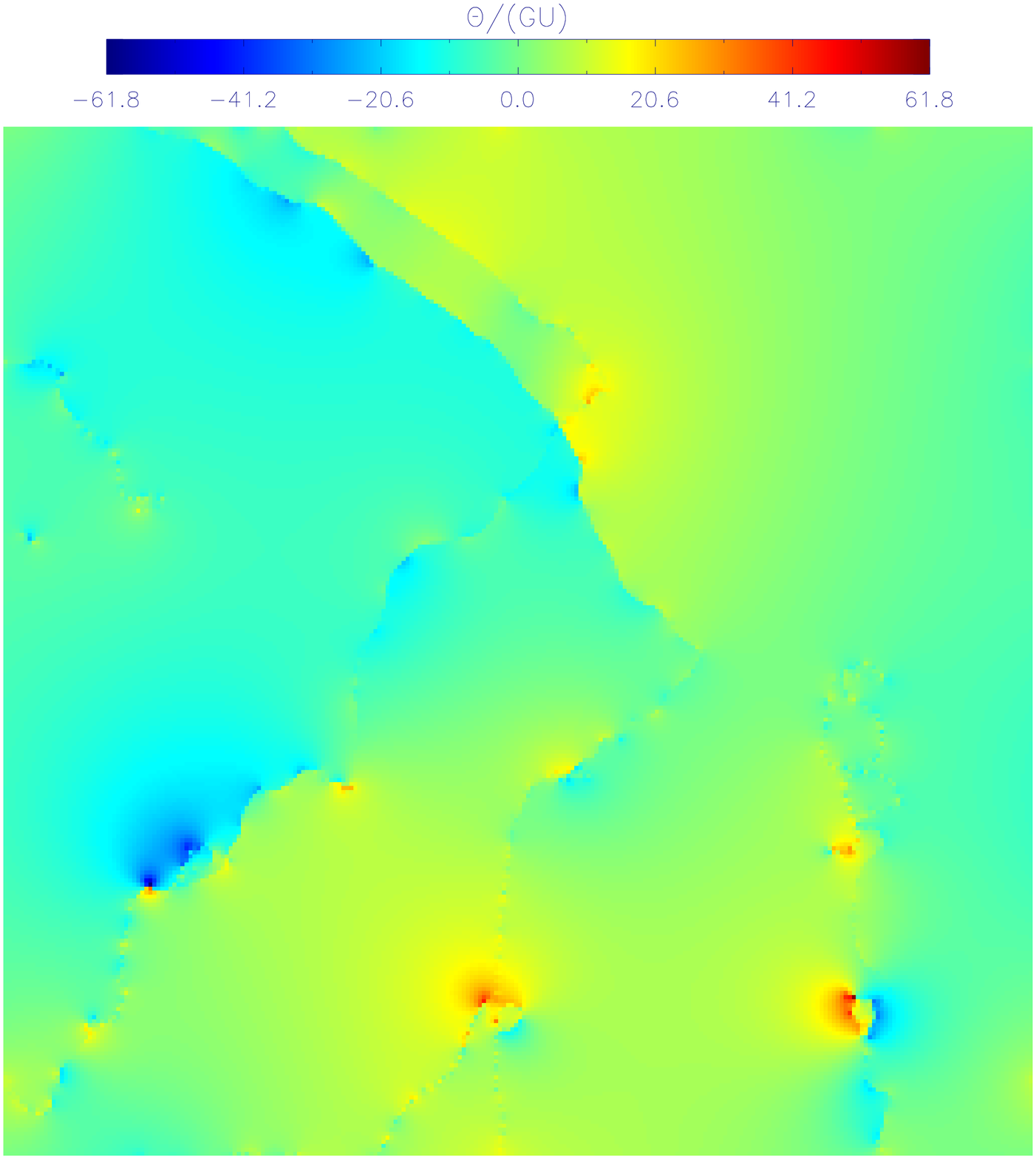}
\includegraphics[width=7.cm]{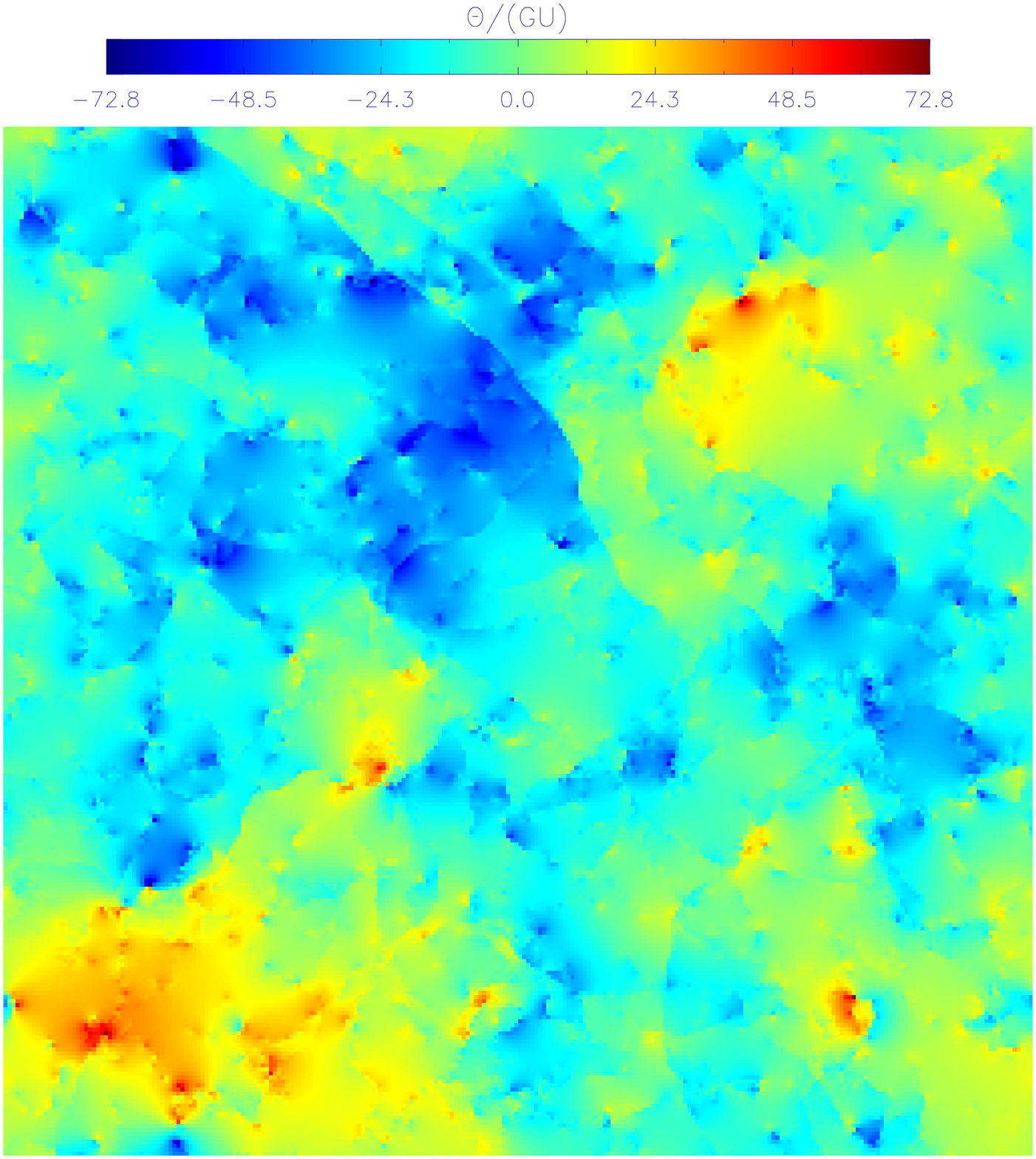}
\includegraphics[width=7.cm]{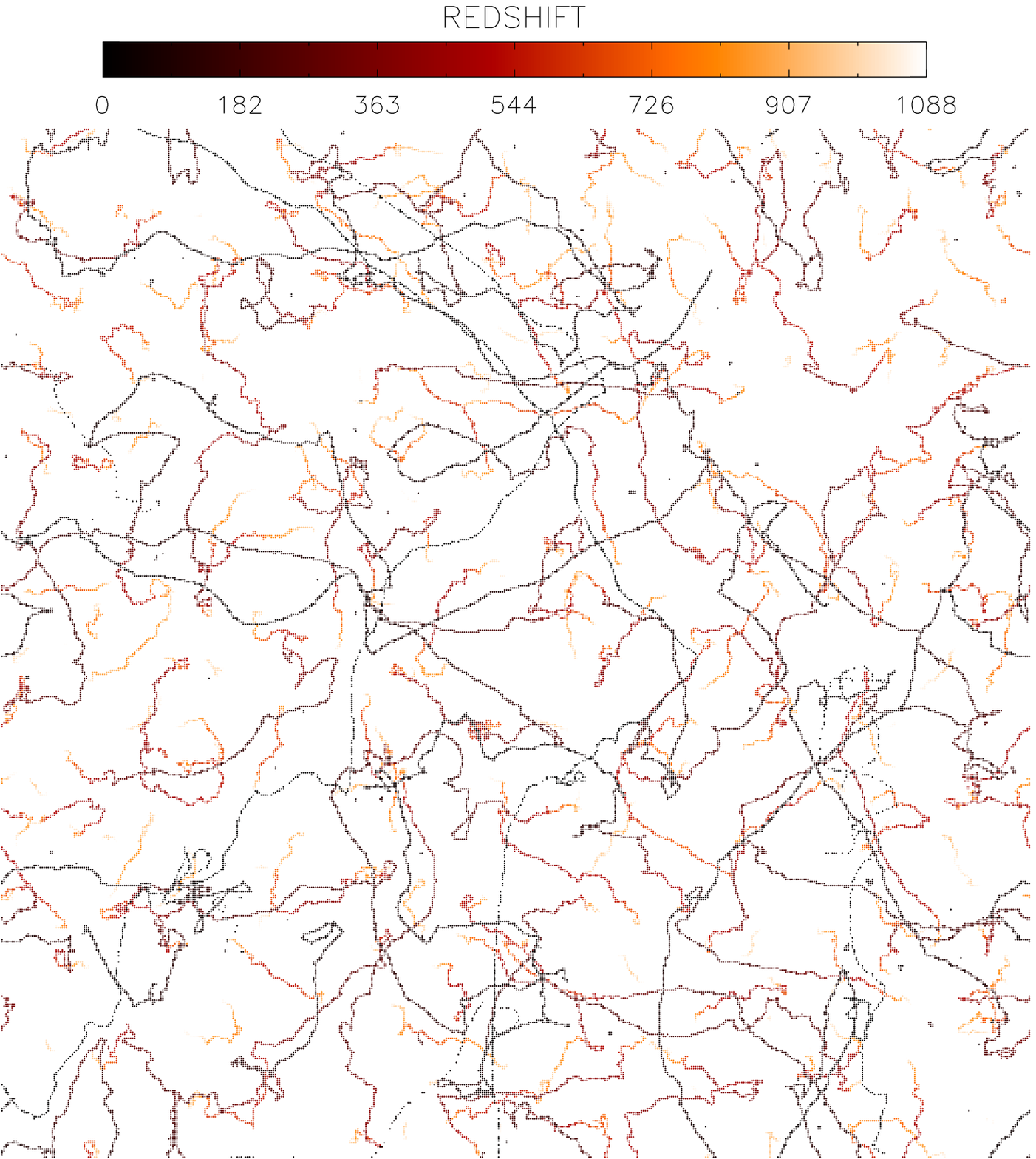}
\caption{String-induced CMB temperature fluctuations on a $7.2^\circ$
  field with a (unrealistic) resolution of $\angres=0.42'$
  ($1024$~pixels).  The upper left image shows the fluctuations
  induced in between the last scattering surface and the redshift
  $z=36$, while the upper right map represents the anisotropies
  produced by strings between $z=36$ and $z=0.3$.  Because of their
  cosmological scaling, most of the long strings intercept our
  past light cone close to the last scattering surface.  The overall
  string-induced fluctuations are plotted in the bottom left panel.
  As can be seen in the bottom right image, the edges in the
  temperature patterns of the other maps can be identified to strings
  intercepting our past light cone.  Note that active regions
  corresponding to string intersection and loop formation events lead
  to the bright spots in these maps.  Some of these spots are
  associated with $\Theta > 80\,\GU$ and saturate the color scale (see
  Sec.~\ref{sec:stats}).} 
\label{fig:dtsplit}
\end{center}
\end{figure*}

The CMB temperature fluctuations induced by strings on a $7.2^\circ$
field of view are shown in Fig.~\ref{fig:dtsplit}. The upper left
(respectively, right) image corresponds to the temperature
fluctuations obtained at the end of the first (respectively, second)
run mentioned above.  The image in the top left panel exhibits more
structures in the temperature patterns than the one in the top right
corner.  This property comes from the cosmological scaling of long
strings which implies that their number inside a Hubble volume remains
approximately constant during the expansion of the Universe.  As a
result, since the Hubble radius grows with cosmic time, the number of
long strings intercepting our past light cone on a given field of view
is higher at high redshift.  It turns out that most of the visible
long strings end up being located close to the last scattering
surface.  Since this implies a small contribution from nearby strings,
we did not perform a third simulation to fill the redshift gap between
$z=0.3$ and $z=0$. 

The overall CMB temperature fluctuations are plotted in the bottom
left panel, while the positions of all the strings intercepting our
past light cone are represented as a function of the redshift of
interception in the bottom right corner.  As can be seen on these
images, the temperature fluctuations appear as a superposition of
discontinuities associated with the long string segments.  The
presence of sharp edges is due to the motion of these segments with
respect to the observer, which induces the typical Doppler shift
associated with the Gott-Kaiser-Stebbins effect~\cite{Gott:1985,
  Kaiser:1984}.  Moreover, a few bright dipolar spots are visible and
come from regions rich in loops and wiggly strings.  These spots can
be associated with the cusps appearing on some isolated loops and with
the kinks sourced in the active regions of string intersections and
loop formation~\cite{Stebbins:1988}.  Although the overall temperature
map is dominated by the long strings pattern, the fluctuations induced
by cusps and kinks may reach very high values.  In fact, since the
bright dipoles come from extremely localized regions, the amplitude of
the temperature jump is smoothed on the scale of a pixel and increases
with increasing image resolution until it reaches its physical value.
As can be seen in Fig.~\ref{fig:dtsplit}, the string-induced
fluctuations observed with a resolution angle of $\angres=0.42'$
($1024$ pixels) clearly exhibit highly non-Gaussian structures.  The
basic statistical properties of these maps are further discussed in
Sec.~\ref{sec:stats}.


\vspace{0.10cm}
\subsection{Robustness}
\label{sec:robust}

\begin{figure*}
\begin{center}
\includegraphics[width=7.cm]{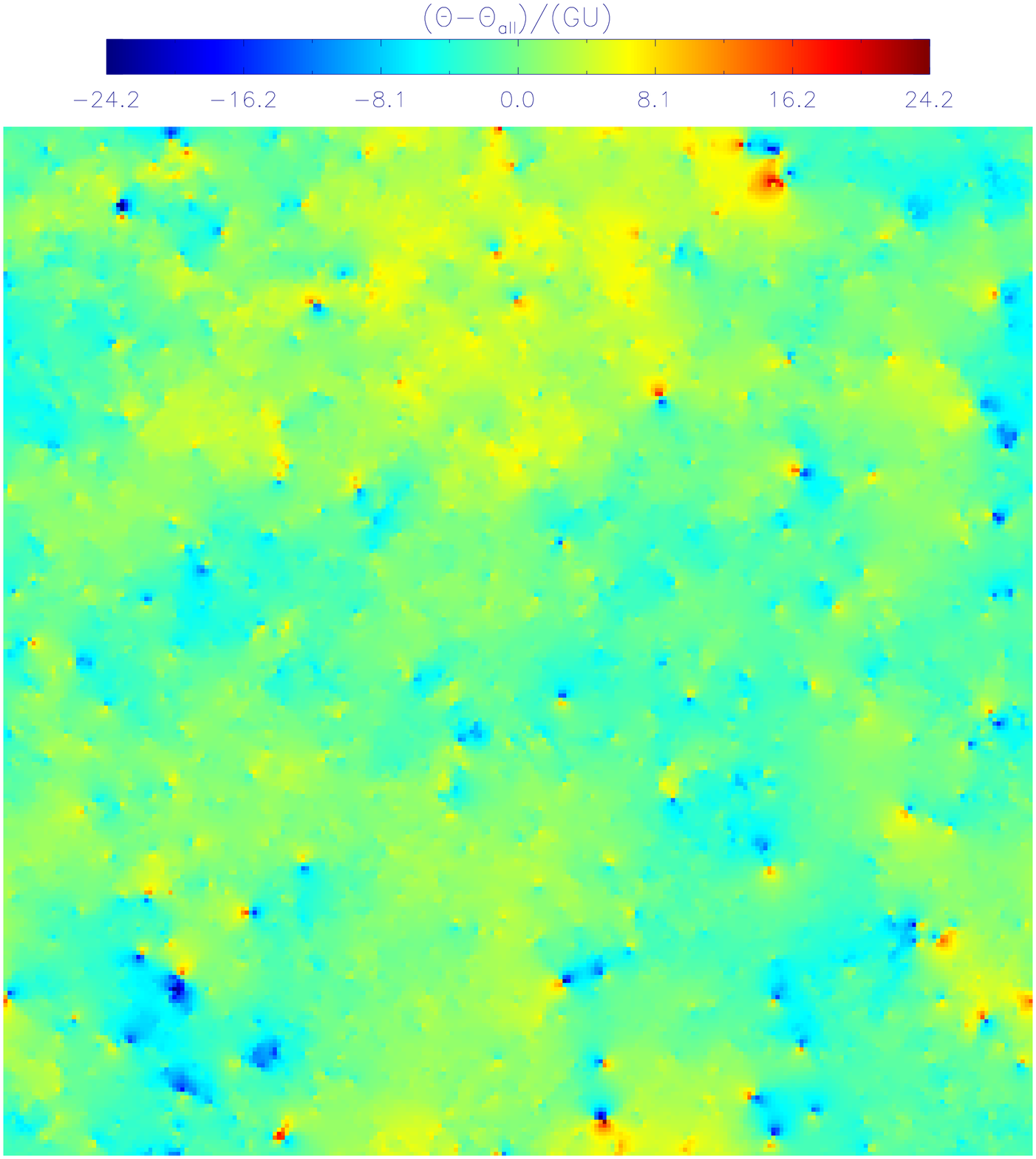}
\includegraphics[width=7.cm]{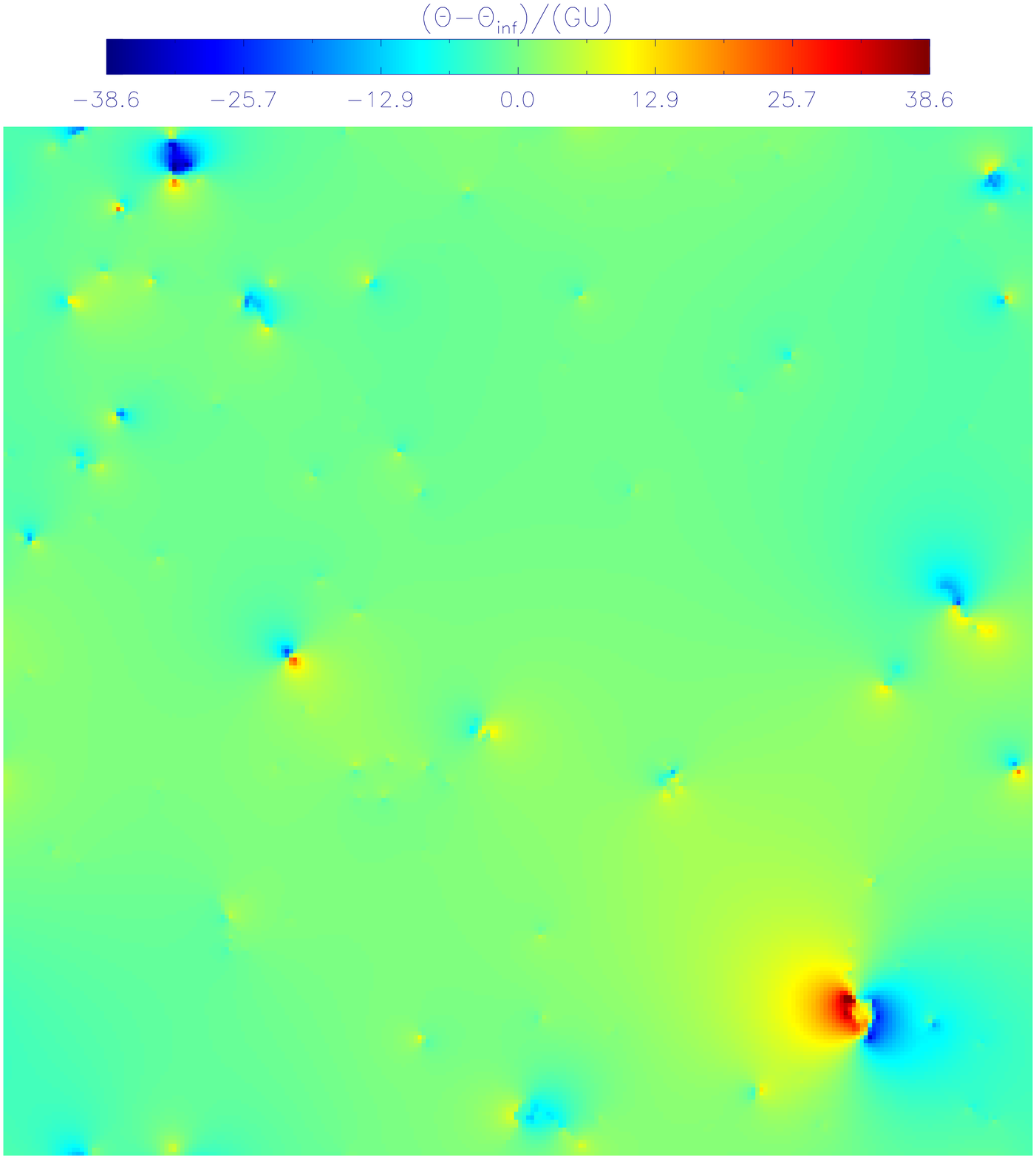}
\caption{Influence of the nonscaling structures on the computed CMB
  temperature maps.  $\Theta_\uall$ (respectively, $\Theta_\uinf$)
  refers to the CMB fluctuations that would be obtained by keeping
  (respectively, removing) all loops present in the numerical
  simulations, including the spurious ones coming from the relaxation
  of the numerical initial conditions.  The left (respectively, right)
  panel shows the deviation induced by the presence (respectively,
  absence) of these structures with respect to the reference
  temperature map shown in the bottom left corner of
  Fig.~\ref{fig:dtsplit}.  We use these maps in Sec.~\ref{sec:stats}
  to estimate the systematic errors induced by the presence of
  nonscaling structures in the cosmic string simulations.} 
\label{fig:dtloop}
\end{center}
\end{figure*}

The CMB temperature fluctuations maps shown in Fig.~\ref{fig:dtsplit}
have been produced using the method described in
Sec.~\ref{sect:csevol} to minimize the influence of the numerical
initial conditions.  The cutoff is adjusted at each redshift to keep
all the loops which have entered their expected stable cosmological
evolution and remove all the smaller loops which may come from the
relaxation of the initial network configuration.  A side effect of
this procedure is to also remove the small loops which may be produced
during the stable cosmological regime.  Depending on the importance of
these small loops on the resulting CMB temperature fluctuations, the
simulated maps we obtain may therefore deviate from the ones that
would be induced by a real string network.  In order to quantify this
deviation, we can compare our results (i.e., the map shown in the
bottom left corner of Fig.~\ref{fig:dtsplit}, thereafter, the
``reference map'') to the maps computed in two extreme cases: when all
the loops are removed from the runs, and when all the loops, even the
smallest ones believed to come from the initial conditions, are kept.
Those results are shown in Fig.~\ref{fig:dtloop}.  As one may expect,
the deviations are most apparent on the smallest scales.  When all
loops are included, the altered map differs from the reference map by
about $10\%$.  The difference is a primarily small-scale noise with a
few bright dipolar spots coming from the tiny loops.  The situation is
similar for the map obtained by removing all the loops.  In that case,
the few dipoles and large-scale structures produced by the loops in
scaling are missing.  Thanks to these ``extreme'' maps, we are now able
to estimate the systematic errors associated with the presence of
nonscaling structure in our numerical simulations.  This is in
particular useful to see how robust are the basic statistical
properties of our maps, which we discuss in Sec.~\ref{sec:stats}. 

\section{Basic statistical properties}
\label{sec:stats}

The CMB temperature patterns are directly related to the properties of
the strings intercepting our past light cone in a particular
realization of the associated network (see Fig.~\ref{fig:dtsplit}).
In order to extract meaningful properties from the string-induced CMB
anisotropies, we generate $84$ statistically independent maps
following the techniques described in Sec.~\ref{sec:simmaps}.  Two
CPU years have been devoted to performing $28$ runs starting from
statistically independent Vachaspati-Vilenkin initial conditions
(on~IBM Power4 P660+ and AMD x86-64 $2\,\GHz$ processors).  Each of
these numerical simulations gives three temperature maps along the
three spatial directions that can be used as either the low or high
redshift contribution.  Finally, the overall temperature anisotropies
map is obtained by combining a low and a high redshift map coming from
two different runs. 


\subsection{One-point distribution function}

\begin{figure*}
\begin{center}
\includegraphics[width=7.cm]{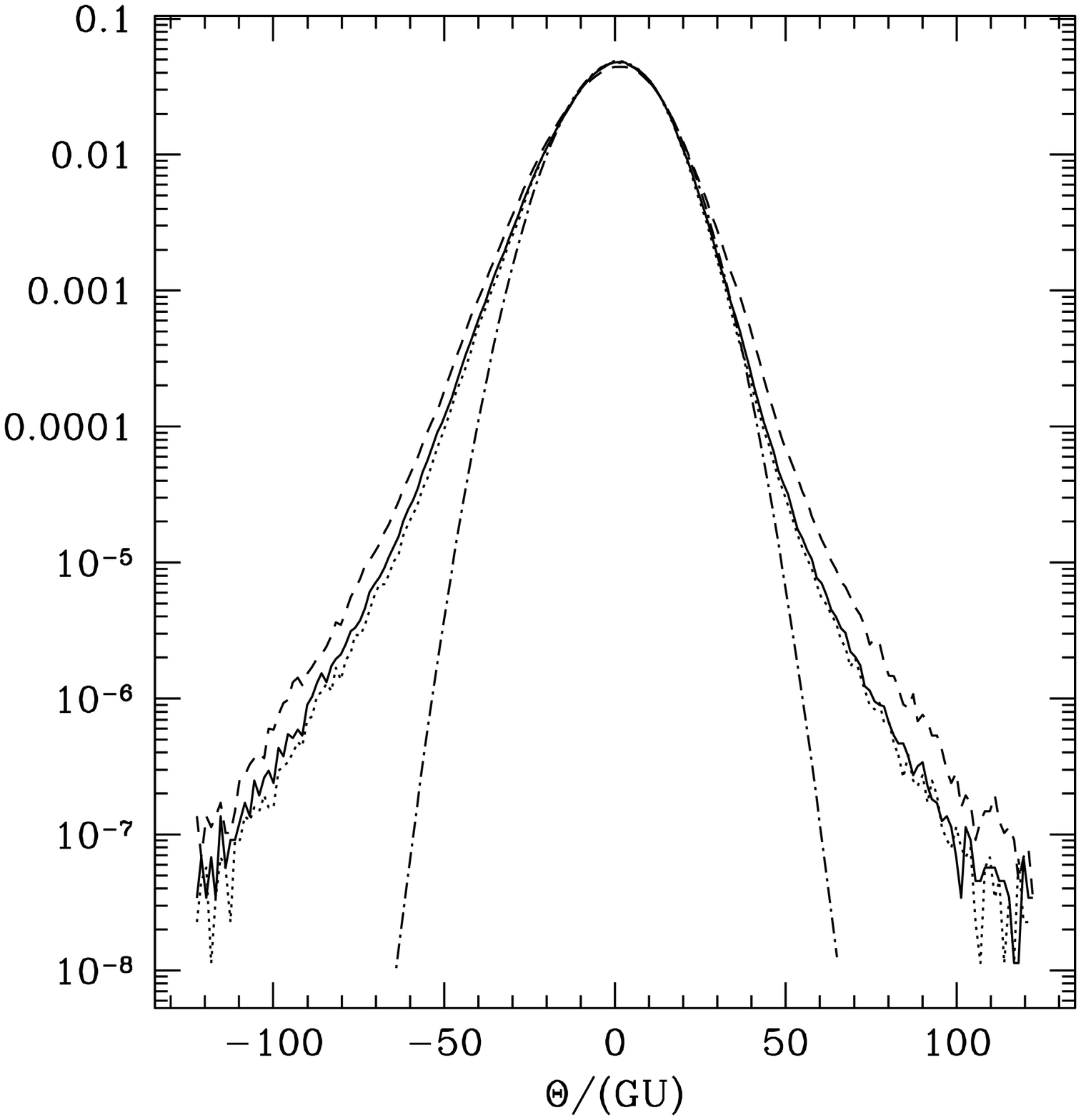}
\includegraphics[width=7.cm]{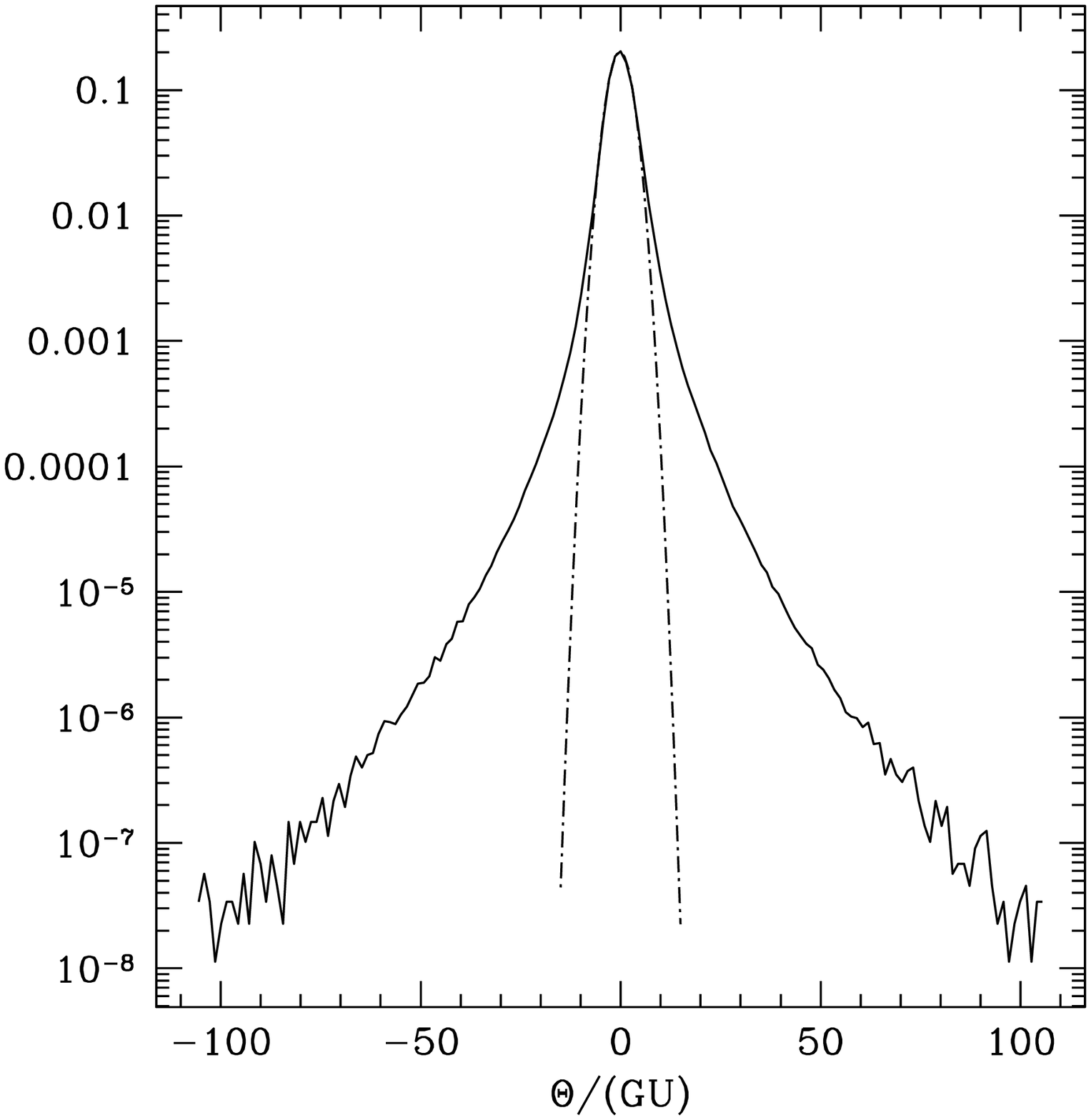}
\caption{The left panel shows the probability distribution function of
  the CMB temperature fluctuations induced by cosmic strings (solid
  line).  The dotted and dashed curves quantify the systematic errors
  coming from the string simulations.  Each of these one-point
  functions is averaged over $84$ independent realizations.  The
  dash-dotted curve represents the best Gaussian fit.  Deviations from
  Gaussianity are clearly apparent in the tails of the distribution,
  as well as from the slight skewness.  The right panel shows the
  probability distribution function of the CMB temperature
  fluctuations that would have been induced by the nonscaling 
  structures, as defined in Sec.~\ref{sec:robust}.  Once again, the
  dash-dotted curve represents the best Gaussian fit.  A slight
  positive skewness may be observed, suggesting that the negative
  skewness observed in the left panel is the result of the loop
  formation~mechanism.} 
\label{fig:1pt}
\end{center}
\end{figure*}

The non-Gaussian nature of the temperature fluctuations induced by
strings produces a very characteristic pattern on the maps shown in
Fig.~\ref{fig:dtsplit}.  A simple quantitative test of this feature is
given by the one-point probability distribution function (PDF) of the
temperature anisotropies.  The normalized PDF has been derived in
three cases: temperature anisotropies generated by the infinite
strings only ($\Theta_\uinf$ map), by the strings and loops in
scaling, and by all the structures ($\Theta_\uall$ map).  The
corresponding PDF's averaged over all the simulated maps are shown in
the left panel of Fig.~\ref{fig:1pt}.  In each case, the PDF
unambiguously deviates from a Gaussian as can be seen by comparison to
the best Gaussian fit to our best PDF estimate (corresponding to the
case where only strings and loops in scaling are taken into account).
It is also worth pointing out that the result is relatively
insensitive to the inclusion of loops, although including nonscaling
structures increases the width of the \nobreak{distribution}.   

More quantitatively, the mean of the distribution is zero to better
than one part in $10^5$, whereas the variance averaged over all $84$
maps is $\left\langle \sigma_\us^2 \right\rangle \simeq
154^{+24}_{-8}\, (\GU)^2\, \Tcmb^2$.  The systematic errors quoted in
this result have been derived from the mean values of the variance
coming from the $\Theta_\uall$ (upper bound) and $\Theta_\uinf$ (lower
bound) maps.  A power-law fit of the tails of the PDF's restricted to
fluctuations with $|\Theta| > 50\, \GU$ shows that they vary as
$\Theta^{-q}$ with $q=8.2\pm 0.2$, which, although steep, decays much
more slowly than a Gaussian.

As can be seen in the shape of the PDF compared to the best Gaussian
fit, we have also measured a slight negative mean sample skewness 
\begin{equation}
\label{eqn:skew}
\langle g_1 \rangle \equiv 
\left<\frac{\overline{(\Theta-\bar{\Theta})^3}}{\sigma^3}\right> 
\simeq -0.24, 
\end{equation}
where $\langle \cdot \rangle$ again refers to the average over all
$84$ maps, whereas a bar corresponds to an average over a single map.
Although the dispersion of $g_1$ between the maps is significant,
there is a clear preference for $g_1<0$ with less than $15$\% of our
maps exhibiting a positive skewness.  This skewness
may come from high correlations between the string velocities and the
shape of the strings along the past light cone.  To confirm this, we
checked that the skewness disappears if the velocity field along the
long strings is randomly redistributed. Since the shape and velocity
field of strings strongly depend on their nonlinear evolution, which
itself relies on intercommutation and loops formation, one may
interpret this effect as a consequence of their nontrivial
interactions.  This interpretation can be further explored by looking
at the PDF of the temperature fluctuations associated with the
``nonscaling'' structures, as defined in Sec.~\ref{sec:robust}. This
PDF is shown in the right panel of Fig.~\ref{fig:1pt} and exhibits a
slightly positive skewness.  Two kinds of structures contribute to
this PDF: the tiny loops generated during the stable cosmological
evolution of the network, and the loops generated by relaxation of the
initial conditions.  It is precisely because of the difficulty of
distinguishing the physical loops from the ones associated with the
relaxation of the initial conditions that we discarded these
structures when generating our maps.  As a result, we stress that
using this PDF as a tool to understand the physics is not
straightforward.  The fact that the PDF shown in the right panel of
Fig.~\ref{fig:1pt} is also skewed supports the existence of the
correlations mentioned above.  However, since the nonscaling
structures are by definition still sensitive to the numerical initial
conditions, the bare value of the skewness given in
Eq.~(\ref{eqn:skew}) should be taken with a grain of salt.  In
particular, one may not exclude a smaller value if all the loops were
evolving in their scaling regime, as it should be for a physical
string network.  To complement this qualitative explanation, it would
be interesting to try to understand this effect quantitatively using
analytical methods, such as the velocity dependent one-scale model of
Refs.~\cite{Martins:1996, Martins:2002}.  Let us finish this
discussion by pointing out that the value of the kurtosis for the PDF
presented in the left panel of Fig.~\ref{fig:1pt} is
\begin{equation}
\langle g_2 \rangle \equiv
\left<\frac{\overline{(\Theta-\bar{\Theta})^4}}{\sigma^4}\right> -3 
\simeq 0.63\,. 
\end{equation}
Gaussian fluctuations, such as the primary anisotropies of
inflationary origin, are entirely described by their one-point
distribution and two-point correlation functions.  As the
string-induced anisotropies are not Gaussian, the angular power
spectrum does not contain all the statistical information associated
with them, but provides a way of estimating their relative
contribution to the CMB anisotropies as a function of angular~scale.


\subsection{Power spectrum}
\label{sec:powspec}

\begin{figure*}
\begin{center}
\includegraphics[width=7.cm]{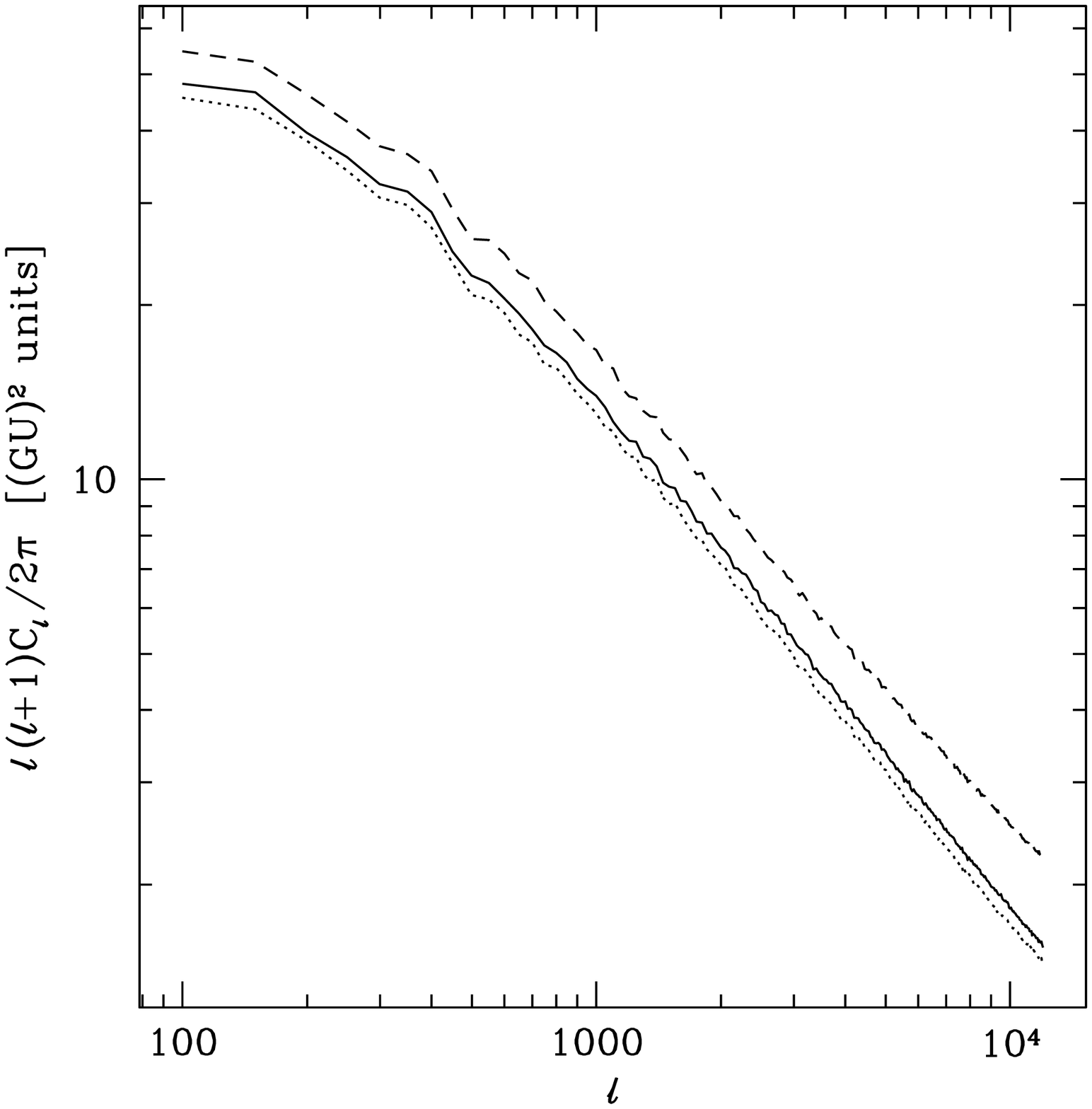}
\includegraphics[width=7.cm]{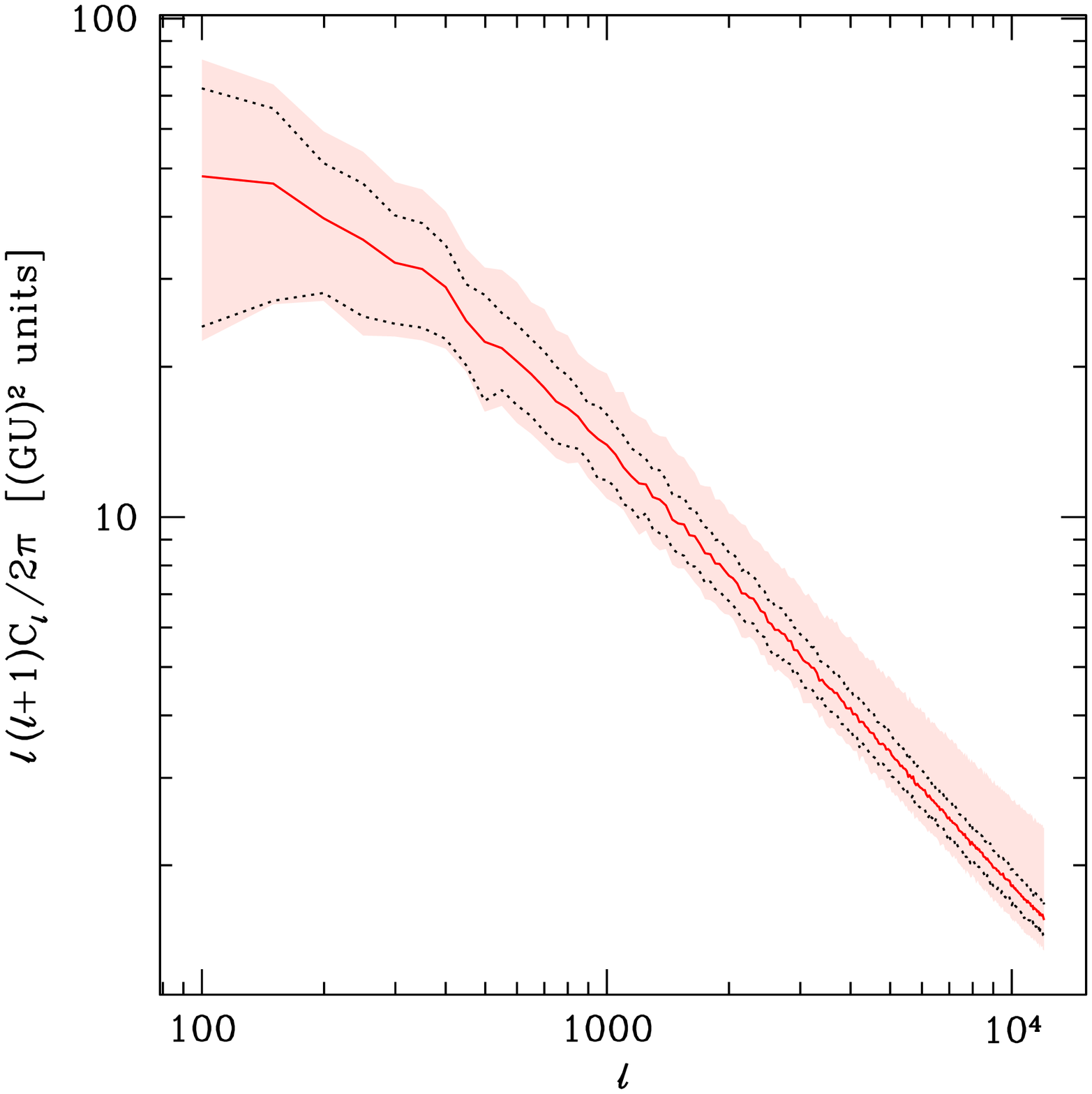}
\caption{Mean (over our $84$ maps) of the angular power spectrum of
  string-induced CMB anisotropies (solid lines).  In the left panel,
  the dashed (respectively, dotted) line is the mean power spectrum
  obtained from string simulations including all loops (respectively,
  infinite strings only).  These two curves give an estimate of the
  systematic error in the results derived from our numerical
  simulations.  In the right panel, the dotted lines define the $\pm
  1$-$\sigma$ statistical error envelope around our best estimate of
  the power spectrum.  The shaded area includes the effect of our
  systematic errors in addition to the statistical ones.} 
\label{fig:strings_spectra}
\end{center}
\end{figure*}

The two-dimensional power spectrum of the string-induced temperature
anisotropies can easily be derived in Fourier space.  Assuming
statistical isotropy, 
\begin{equation}
\powspec(\uwav) = \left|\dfrac{1}{\GU} \int \Theta(\angx,\angy)\,
  \ue^{-\mathrm{i}\,\vect{\uwav} \cdot \vect{\ang}}\, \ud^2\vect{\ang}
  \right|^2,
\end{equation}
where $\vect{\ang}\equiv(\angx,\angy)$ denotes angular coordinates and
$\vect{\uwav}$ the two-dimensional wave vector.  In the small-angle
approximation, this is also the angular power spectrum up to a
redefinition of the wave number~\cite{White:1999}.  We have 
\begin{equation}
\dfrac{\ell(\ell+1)}{4 \pi^2}\,C_\ell \simeq \uwav^2 \powspec(\uwav),
\end{equation}
with $\uwav = \ell/(2\pi)$, which we use to produce
Fig.~\ref{fig:strings_spectra}.  For a field of view of $7.2^\circ$
and a resolution of $0.42'$, the multipole moments run from
$\ell\simeq 10^2$ to $10^4$.  As detailed in Sec.~\ref{sec:simmaps},
we only take the string-induced ISW effect into account when computing
the CMB temperature fluctuations from strings.  As a result, the
spectra we present in this section are only accurate at small angular
scales, typically for multipoles higher than a few hundred.  The left
panel of Fig.~\ref{fig:strings_spectra} shows three spectra averaged
over our $84$ maps.  The central one corresponds to our best estimate,
whereas the top and bottom spectra have been, respectively, obtained
from the string-induced temperature anisotropies $\Theta_\uall$ and
$\Theta_\uinf$ and trace the systematic error in the power spectrum
estimation (see Sec.~\ref{sec:robust}).  In the right panel, we show
our best estimate of the power spectrum, as well as of the associated
systematic and statistical errors combined.  We define this
combination as the smallest area around our best estimate containing
all the $\pm 1$-$\sigma$ statistical error envelopes of the spectra
presented in the left panel.

Although the uncertainties induced by the presence of nonscaling
structures in the simulations affect the overall normalization of the
spectrum, its global shape remains the same.  Using a power law fit to
the small-angle tail of the spectrum, 
\begin{equation}
\label{eq:clsfit}
\ell(\ell+1)\, C_\ell \underset{\ell \gg 1} \propto \ell^{-p}
  \quad \mathrm{with} \quad
  p = 0.889^{+0.001}_{-0.090}\,,
\end{equation}
where we only quote systematic errors estimated from the extreme
power spectra shown in the left panel of
Fig.~\ref{fig:strings_spectra}.  Let us point out the fact that the
overall amplitude, in units of $\GU$, of our best estimate spectrum
appears to be compatible with the one recently obtained for Abelian
strings at large angular scales by the UTC method~\cite{Bevis:2007}.
At multipole $\ell=1000$, we have 
$\ell(\ell+1)\,C_\ell/(2\pi)\simeq\nobreak 14\,(\GU)^2$.  It is nice to see
these two techniques agree over the range of multipoles where they are
both valid approximations, but it is not trivial that this should be
the case.  In particular, numerical simulations of Abelian string
networks do not exhibit the typical loop formation events observed for
Nambu-Goto strings~\cite{Vincent:1998,Moore:2002}.  However, the
scaling regime of long strings is reached in both Abelian and
Nambu-Goto simulations.  As already discussed, the typical distance
between long strings in scaling at the surface of last scattering is a
fraction, typically a fifth~\cite{Ringeval:2007}, of the Hubble radius
at that time.  As a result, even the long strings are contributing to
the CMB anisotropies down to moderately small angles.  It is therefore
not totally surprising that both spectra are similar at intermediate
angular scales ($1000 \lesssim \ell \lesssim 3000$).  We do not try to
compare our results to those presented in~\cite{Bevis:2007} at larger
angular scales since our method relies on the flat-sky approximation,
which breaks down for multipoles smaller than a few hundred.

As can be seen in Fig.~\ref{fig:strings_spectra}, some extra power
shows up at very small scales for the $C_\ell$'s coming from the
numerical simulations including the nonscaling loops
($\Theta_\uall$).  The same effect was already visible on the
corresponding temperature map plotted in Fig.~\ref{fig:dtloop} (left
panel) and suggests that nonscaling structures start to have
significant effects at very small scales, for
$\ell \gtrsim 10^4$.  The very small-scale effect of loops (be they in
scaling or not) is also what explains the asymmetric errors appearing
in Eq.~(\ref{eq:clsfit}).  As expected from the presence of cosmic
strings all along our past light cone, the power spectrum decays quite
slowly at small angular scales, and much slower than the exponential
Silk damping experienced by the CMB anisotropies of inflationary
origin.  As a result, a contribution of cosmic strings to the angular
power spectrum could be negligible at low multipoles and yet dominate
the primary anisotropies for large values of $\ell$.  We discuss the
observability of string-induced anisotropies at small scales
in more detail~in~Sec.~\ref{sec:obs}.


\subsection{Gradient magnitude as a string tracer}

Although the one- and two-point functions already exhibit some
characteristic features of string-induced CMB anisotropies, they
remain weakly sensitive to most of the highly non-Gaussian features
one can see in Fig.~\ref{fig:dtsplit}.  More sophisticated statistical
tools have been developed to look for patterns in the CMB.  These
techniques involve the use of Minkowski functionals, the skeleton,
wavelet analysis, or temperature step finders, among
others~\cite{Mecke:1994, Hobson:1999, Komatsu:2003, Jeong:2007,
  Sousbie:2007}.  The corresponding estimators are usually applied
directly on CMB temperature maps.  However, the deviations from
Gaussianity appearing in the PDF shown in the left panel of
Fig.~\ref{fig:1pt} are due to two main effects, namely, the presence
of rare high temperature events coming from localized string and loop
regions with kinks and cusps, and the linelike (and fractal) shape of
the temperature steps induced by moving strings in the microwave sky,
which also explains the power-law behavior of the power spectrum at
small angular scales.  One may therefore expect such estimators to be
more efficient when directly applied to a string map, such as the one
shown in the bottom right panel of Fig.~\ref{fig:dtsplit}.  Of course,
such a map cannot be directly observed.  But since strings appear as
discontinuities in the temperature maps, the temperature gradient
would be singular at the string location, enabling us to produce a map
similar to the one shown in the bottom right panel of
Fig.~\ref{fig:dtsplit}, though with no redshift information.  This map
could then be used with the techniques mentioned above to get more
statistical information on the properties of networks of strings. 

\begin{figure}
\begin{center}
\includegraphics[width=7.cm]{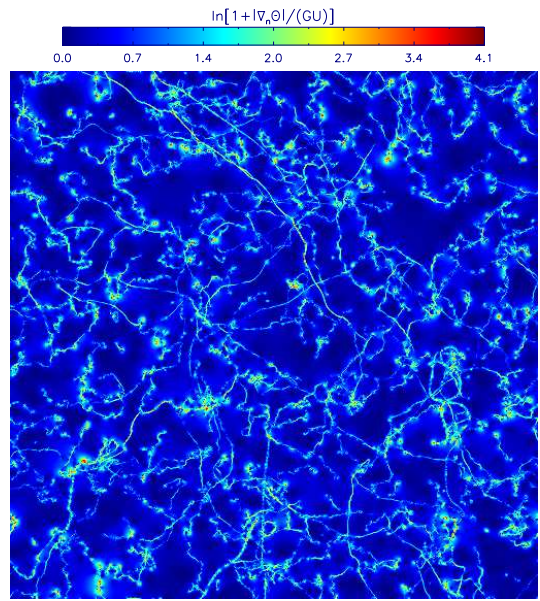}
\caption{Normalized gradient magnitude of the string-induced
  temperature anisotropies shown in Fig.~\ref{fig:dtsplit} (bottom
  left panel).  A logarithmic scale has been used to enhance the
  contrast by preventing the bright spots from saturating the
  color scale.  Such a map reproduces the string path on our past
  light cone and enhances the active string regions (see
  Fig.~\ref{fig:dtsplit}).}
\label{fig:gradmag}
\end{center}
\end{figure}

Directional gradients were first discussed in Ref.~\cite{Gott:1990} to
analyze the topology of string-induced temperature anisotropies.  In
order to conserve isotropy, it is convenient to consider the gradient
magnitude $|\nabla \Theta|$ of the temperature anisotropies defined by 
\begin{equation}
\label{eq:gradmag}
\left| \nabla \Theta \right| \equiv \sqrt{
  \left(\dfrac{\ud \Theta}{\ud \angx}\right)^2 +
  \left(\dfrac{\ud \Theta}{\ud \angy}\right)^2}\, ,
\end{equation}
where $\angx$ and $\angy$ are the horizontal and vertical angular
coordinates.  This definition makes it clear that for a finite
temperature step, let us say $\Theta(\angx,\angy) =  
\Theta_\zero\,\heaviside{\angx-\angx_\zero}$, $\mathrm{H}$ being the
Heaviside function, the resulting gradient magnitude is a Dirac
distribution at the string location.  On a pixelized map, the maximal
amplitude of $|\nabla \Theta|$ is therefore given by the size of each
pixel.  To take this resolution effect into account, we instead look
at the normalized gradient magnitude $|\nablanum \Theta|/\GU$, defined
by 
\begin{equation}
\label{eq:gradnum}
\left|\nablanum\Theta\right| \equiv \left|\nabla\Theta\right|\angres,
\end{equation}
of the temperature map shown in the bottom left panel of
Fig.~\ref{fig:dtsplit}.  The resulting map is shown in
Fig.~\ref{fig:gradmag}.  Note that we enhanced the contrast by using a
logarithmic scale to prevent the bright dipoles from saturating the
color scale.  As can be checked by comparing Fig.~\ref{fig:gradmag} to
the bottom right panel of Fig.~\ref{fig:dtsplit}, the gradient map
reproduces the string path on our past light cone.  In addition, the
magnitude of the gradient, once normalized to the image resolution,
encodes the transverse string velocity, which renders kinks and cusps
clearly visible.  But since all the results presented in this section
have been derived for strings alone, and with a high and quite
unrealistic angular resolution, one may question their relevance in
view of the current experiments and data.  In the next section, we
address this issue by combining the string-induced anisotropies with
the standard primary and expected secondary CMB signals for a typical
arcminute-resolution CMB~experiment.


\section{Observing cosmic strings}
\label{sec:obs}

CMB experiments have so far been able to map the microwave sky down to
scales corresponding to multipoles of order $2 \times 10^3$, WMAP
providing cosmic variance limited measurements up to
$\ell \sim 400$~\cite{Spergel:2007}.  At these scales, the primary CMB
dominates over all secondary effects by more than an order of
magnitude.  However, upcoming experiments will produce CMB maps with
arcminute resolution, at which point secondary effects become the
dominant source of
fluctuations~\cite{Ami:2006,Ruhl:2004,Kosowsky:2006}.  It is therefore
crucial to take them into account to produce reasonably realistic maps
of what the CMB would look like at small angular scales in the
presence of strings.  Throughout this section, we consider a flat
power-law $\Lambda$CDM cosmology with $h=0.70$,
$\Omega_\Lambda = 0.73$, $\Omega_\mathrm{b} = 0.045$,
$n_\mathrm{s} = 0.95$, $\Omega_\mathrm{m} = 0.27$, $\sigma_8 = 0.78$
and $\tau = 0.073$.  The hydrogen fraction is set to $0.76$.


\vspace{-0.2cm}
\subsection{Contribution of secondary anisotropies}
\label{sec:seccontrib}

\begin{figure*}
\begin{center}
\includegraphics[width=7.cm]{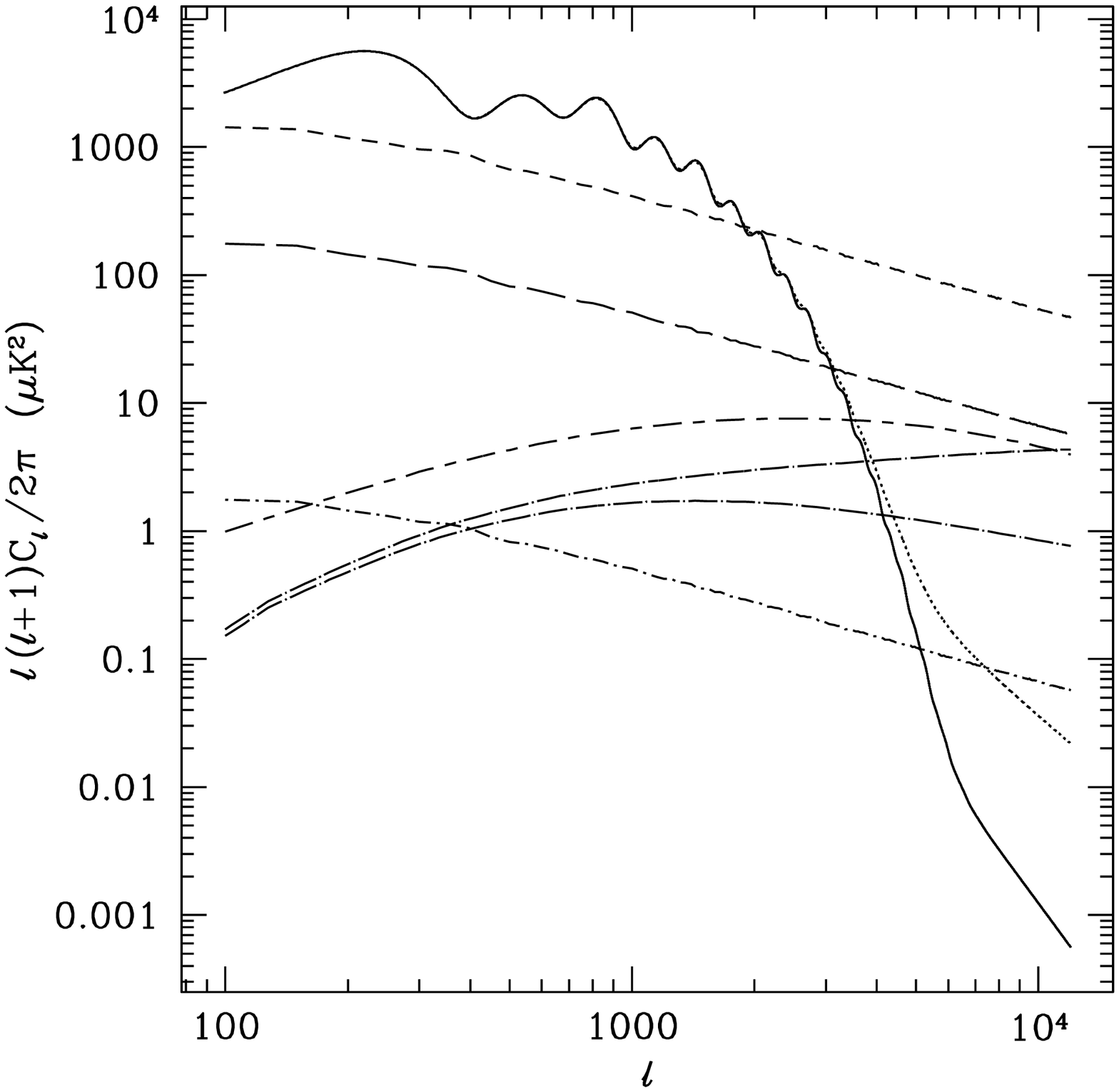}
\includegraphics[width=7.cm]{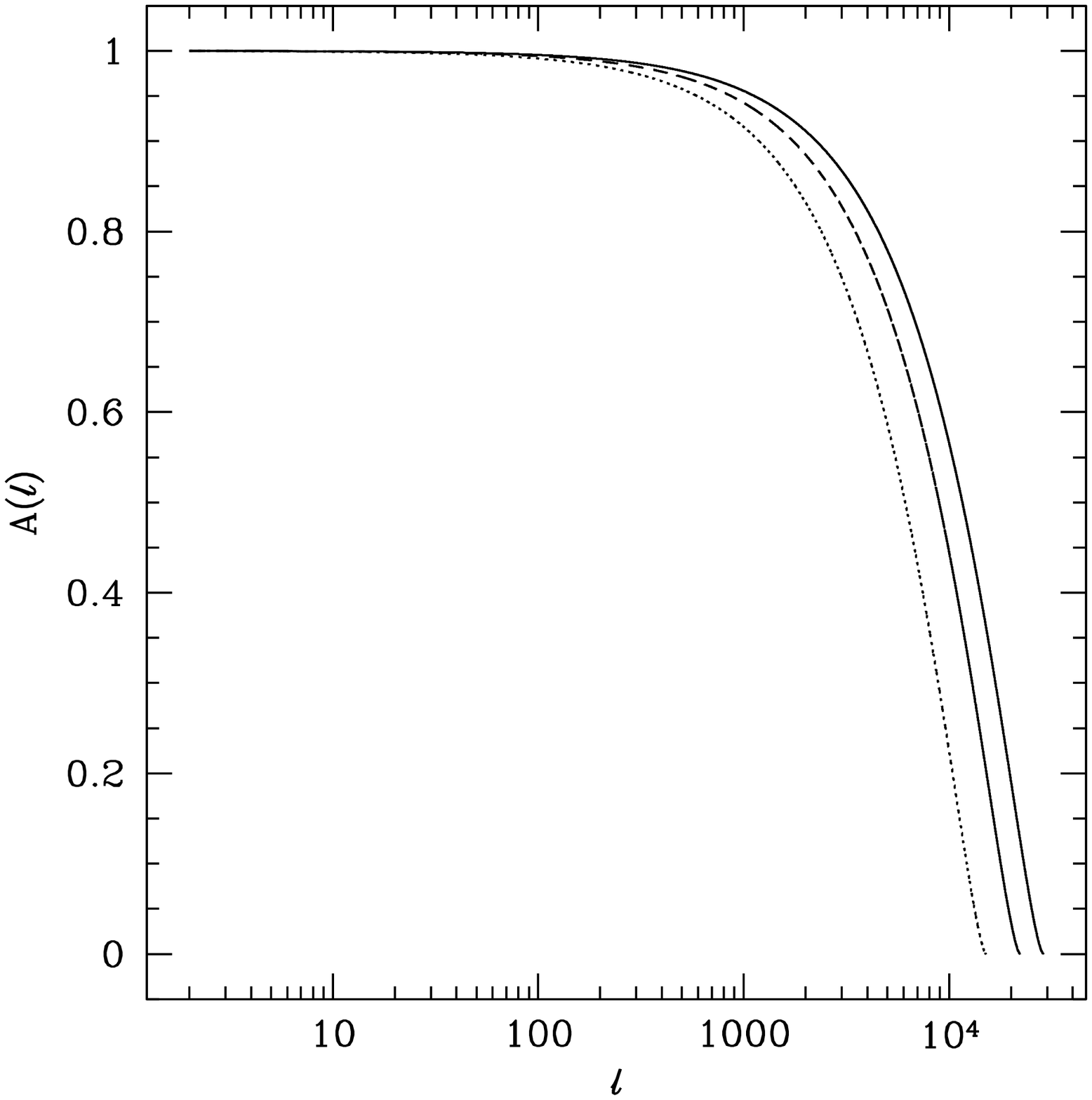}
\caption{(Left) Angular power spectrum of the primary CMB anisotropies
  (solid line), the lensed CMB (dotted line), the thermal
  Sunyaev-Zel'dovich effect (long dashed-short dashed line), the
  Ostriker-Vishniac effect (lower long dash-dotted line), and the
  nonlinear kinetic Sunyaev-Zel'dovich effect (upper long dash-dotted
  line) in a flat power-law $\Lambda$CDM universe (see text).  The
  fluctuations generated by cosmic strings are shown for three values
  of the string tension: $\GU = 2 \times 10^{-6}$ (short dashed line),
  $7 \times 10^{-7}$ (long dashed line), and $7\times 10^{-8}$ (short
  dash-dotted line). (Right) Model of the primary ACT beam in
  multipole space at $279$~GHz (solid line), $215$~GHz (dashed line)
  and $147$~GHz (dotted line).  The normalization is chosen such that
  the beam is unity at $\ell = 0$.  As expected for a
  diffraction-limited beam, the cutoff multipole increases with
  increasing frequency.}
\label{fig:cls_all}
\end{center}
\end{figure*}

In the flat-sky limit, provided that the angular scale be small enough
for the two-halo term to be negligible, the thermal Sunyaev-Zel'dovich
effect (tSZ) angular power spectrum can be computed in the context of
the halo formalism and is then given by~\cite{Sunyaev:1970,
  Sunyaev:1972, Komatsu:2002}
\begin{equation}
\label{eq:tSZ}
\begin{aligned}
C_\ell = & 
  \left(x\,\frac{\mathrm{e}^x + 1}{\mathrm{e}^x - 1} - 4\right)^2
  \int_{0}^{z_{\umax}} \ud z\, \frac{\ud V(z)}{\ud z}\\[1mm]
       &\times \int_{M_{\umin}}^{M_{\umax}} \ud M\,
  \frac{\ud n(M,z)}{\ud M} \left|\tilde{y}_\ell(M,z)\right|^2,
\end{aligned}
\end{equation}
where $x=h\nu/k_\mathrm{B} \Tcmb$, $V(z)$ is the comoving volume at
redshift $z$ per steradian, $\ud n/\ud M$ is the comoving number
density of dark matter halos of mass $M$ at redshift $z$, and 
$\tilde{y}_\ell(M,z)$ is the Fourier transform of the projected
Compton $y$ parameter.  To compute $\mathrm{d}n(M,z)/\mathrm{d}M$, we
use the Jenkins mass function~\cite{Jenkins:2001}, and the linear
matter power spectrum produced by \texttt{CAMB}~\cite{Lewis:2000}.
$\tilde{y}_\ell(M,z)$ is calculated as in Ref.~\cite{Komatsu:2002},
the tSZ profile being integrated over a maximum of two virial radii.
At the angular scales of interest, the values of $z_{\umax}$,
$M_{\umin}$, and $M_{\umax}$ have been set according to
Ref.~\cite{Komatsu:2002} and are accurate enough to ensure the
convergence of the integrals.  However, there are several theoretical
uncertainties in the calculation of the mass function and the Compton
$y$ parameter, leading to a typical uncertainty of a factor of 2 on
the overall amplitude of the tSZ spectrum.  Moreover, as the two-halo
term is neglected, the current approach holds only for
$\ell > \ell_{\umax}\simeq 300$~\cite{Komatsu:1999}.  This limitation
is unimportant for our study as the primary CMB dominates over all
other secondary effects by more than 3 orders of magnitude for
multipoles smaller than $\ell_{\umax}$.

The kinetic Sunyaev-Zel'dovich effect (kSZ) has been computed exactly
in the linear regime, where it is also known as the Ostriker-Vishniac
(OV) effect~\cite{Sunyaev:1980, Ostriker:1986,
  Vishniac:1987}. Although no exact result exists in the nonlinear
regime, several analytical models have been proposed to describe this
component~\cite{Hu:2000, Ma:2002, Zhang:2004}.  Here, we use the model
of Ref.~\cite{Zhang:2004} with, as input, the same linear matter power
spectrum generated by $\CAMB$ as for the tSZ.  Although the nonlinear
corrections are unimportant at large scales, they become significant
in the small-angle limit.  Within our cosmological model, assuming
instantaneous reionization, the nonlinear component of the kSZ has
the same amplitude as the linear one at $\ell \sim 2500$.

Several other secondary effects can produce fluctuations in the CMB
temperature, among which are gravitational lensing, the ISW and
Rees-Sciama effects linked to the standard cosmological fluids, patchy
reionization, and point sources.  The corrections to the primary
anisotropies due to gravitational lensing have been included and
computed with $\CAMB$.  Though they become important relative to the
primary anisotropies for $\ell \geq 2000$, the Sunyaev-Zel'dovich
effects are always dominant at these scales.  The ISW and Rees-Sciama
effects linked to the standard cosmological fluids are never dominant
secondary effects for $\ell \geq 100$~\cite{Hu:2002}, and we therefore
ignore them.  It has been shown that patchy reionization can make a
significant contribution to the nonlinear kSZ, though an agreement on
how big this contribution is is still to be found~\cite{Santos:2003,
  McQuinn:2005, Iliev:2007}.  Given the uncertainties involved in this
effect and the fact that, even if the signal coming from patchy
reionization dominates the nonlinear kSZ, it is still at least an
order of magnitude smaller than the tSZ at all
scales~\cite{Iliev:2007}, we do not include it in our analysis.
Finally, the question of how one should deal with point sources is
probably the trickiest one.  Modeling point sources (be they radio or
infrared ones) is indeed an arduous task leading to quite uncertain
models~\cite{Huffenberger:2005}, which leads us to omit them in the
present analysis.  This is a clear limitation of our study that should
be kept in mind when trying to use the results presented here to
predict how well we could be able to do as far as detecting cosmic
strings is concerned.  Although complementary observations at other
wavelengths will help in removing the signal from the brightest point
sources~\cite{Kosowsky:2006}, our ability to remove the remaining
contamination will limit the possibility of detecting the signature of
cosmic strings in the microwave sky.  This should be yet another
motivation for  trying to improve our understanding of point sources
and develop efficient techniques to remove them without losing
interesting string-induced features. 

The angular power spectra of all these effects, namely, lensed primary
anisotropies, tSZ, OV, and nonlinear kSZ, are shown in the left panel of
Fig.~\ref{fig:cls_all}, along with various spectra of the anisotropies
induced by cosmic strings.  From top to bottom, the string power
spectra have been generated with $\GU=2\times 10^{-6}$, $7\times
10^{-7}$, and $7\times 10^{-8}$.  The tension $2\times 10^{-6}$ is the
one required to normalize the large-scale string-induced power spectrum
presented in Ref.~\cite{Bevis:2007} to the WMAP three-year data at
$\ell = 10$.  Such a high value of the string tension is already ruled
out at more than $3$-$\sigma$~\cite{Slosar:2006}, but is commonly used
as a reference.  $GU = 7 \times 10^{-7}$ is the $95\%$ confidence
level upper limit set in Ref.~\cite{Hindmarsh:2007} with the power
spectrum presented in Ref.~\cite{Bevis:2007}.  Since the amplitude of
our string spectrum at $\ell\simeq 1000$ matches the one derived in
this reference (see Sec.~\ref{sec:powspec}), we use
$\GU=7\times 10^{-7}$ to produce the maps shown in
Figs.~\ref{fig:mixed_d7EM7} and~\ref{fig:mixed_ob7EM7}.  Let us point
out the fact that the associated string spectrum becomes marginally
dominant over all primary and secondary anisotropies for multipoles
larger than typically $3\times 10^3$.  However, for smaller tensions,
the dependence of the amplitude of the string-induced power spectrum
on $(GU)^2$ rapidly brings the latter below the spectra of all the
secondary anisotropies we consider for $\ell$ above a few hundred.
This is the case, e.g., for $GU = 7\times 10^{-8}$.  Because of this
strong dependence on $GU$, the factor of 2 uncertainty on the
amplitude of the tSZ is relatively unimportant for our study.

Combining all these results, we can now generate simulated maps of the
temperature fluctuations produced by the corresponding primary and
secondary effects and add them to the string-induced anisotropies.
However, to more realistically predict what arcminute-resolution CMB
observations would look like in the presence of strings, we also need
to convolve the resulting maps with a model for the beam of such an
experiment.


\vspace{-0.2cm}
\subsection{Instrument beam}
\label{sec:beam}

\begin{figure*}
\begin{center}
\includegraphics[width=7.cm]{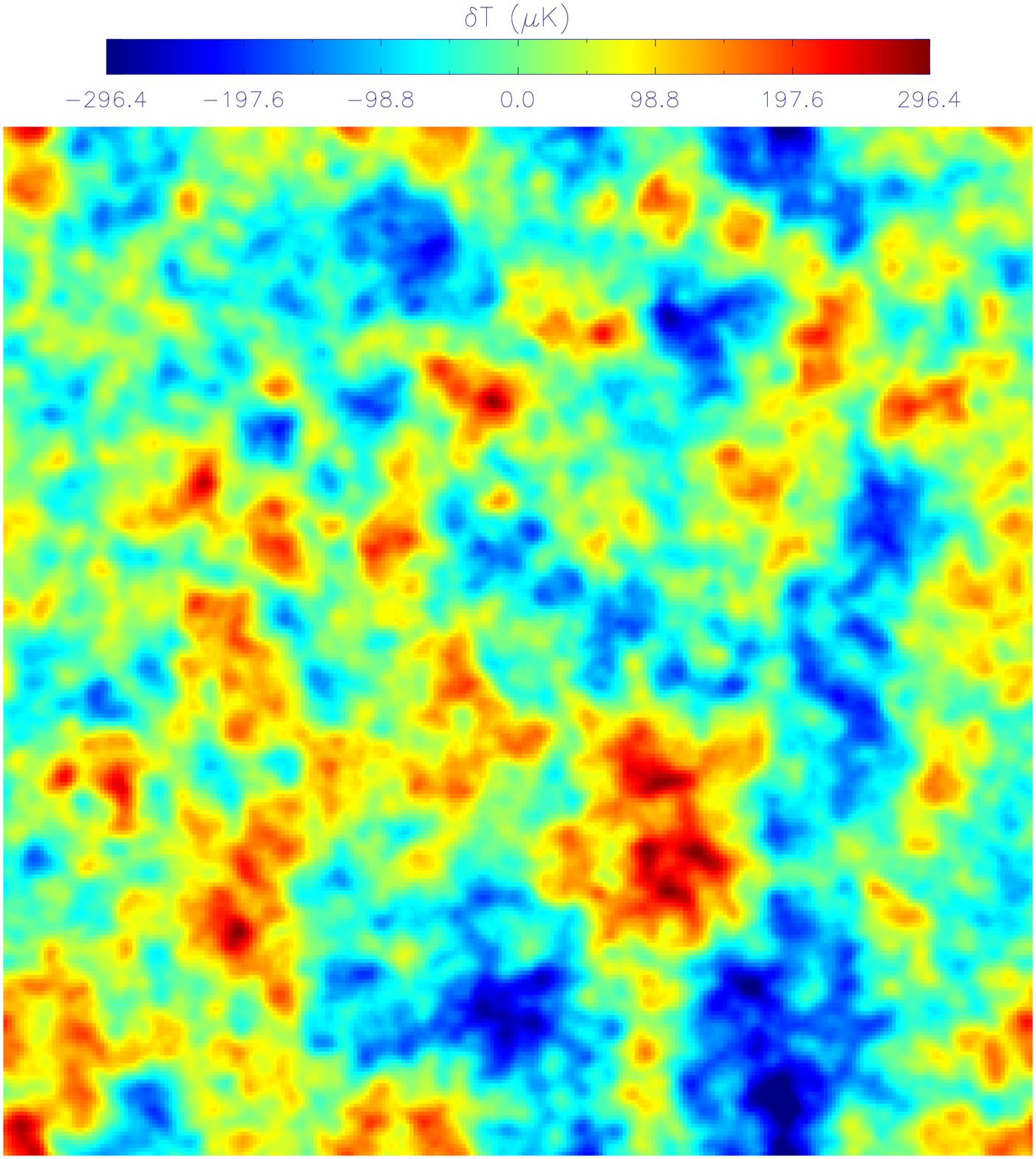}
\includegraphics[width=7.cm]{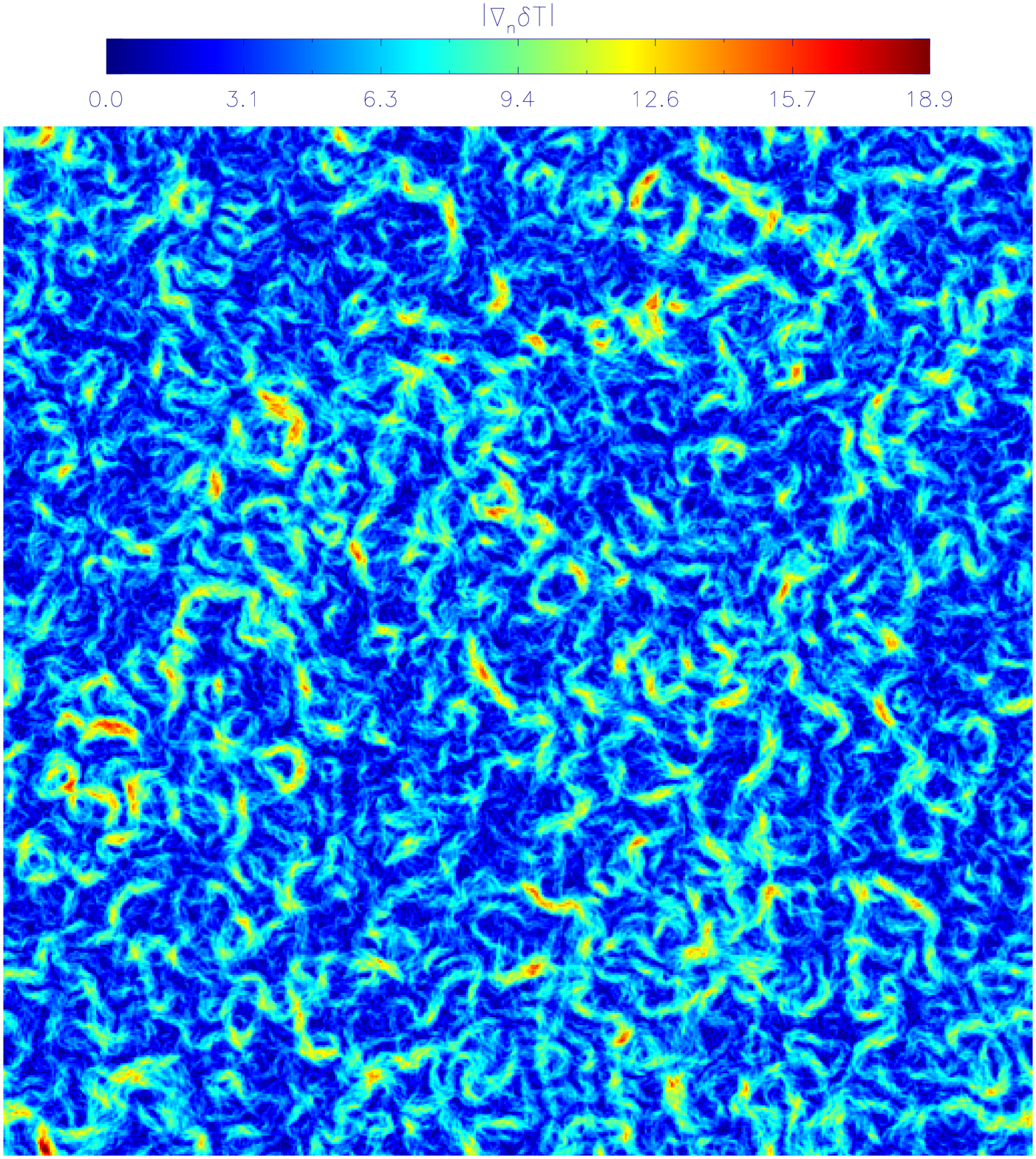}
\caption{CMB temperature anisotropies, including the lensed primary
  fluctuations predicted by our fiducial $\Lambda$CDM model (see
  text), as well as the tSZ, OV, and nonlinear kSZ effects, for a
  $7.2^\circ$ field of view at $147\,\GHz$ (left), and corresponding
  normalized temperature gradient magnitude map (right).  These maps
  take into account the effect of the primary ACT beam as derived in
  Sec.~\ref{sec:beam}.  The effect of strings is \emph{not} included.} 
\label{fig:mixed_0}
\end{center}
\end{figure*}

Several arcminute-resolution CMB experiments are already underway,
covering frequencies from $15$~GHz~\cite{Ami:2006} all the way to
$345$~GHz~\cite{Ruhl:2004}.  In the remaining of this paper, we will
use a model based on the ACT specifications to include realistic
resolution effects in our maps.  ACT is a 6 m telescope that will map
the microwave sky at $147$, $215$, and
$279$~GHz~\cite{Kosowsky:2006}.  The actual data collected in such an
experiment is not directly the temperature on the sky, but the result
$\mathcal{V}$ of the convolution of the latter with the instrument
beam such that~\cite{White:1999}
\begin{equation}
\label{eqn:convol_beam}
\mathcal{V(\vect{k})} = \frac{\partial B_\nu (T)}{\partial T} \, \Tcmb
  \int \Theta(\vect{r}) \, A(\vect{r}) \,
  \mathrm{e}^{-\mathrm{i}\,\vect{k}\cdot\vect{r}} \, \ud ^2 \vect{r},
\end{equation}
where $B_\nu (T)$ is the Planck function, $\vect{r}$ the coordinates
of a point on the telescope, and $A(\vect{r})$ the primary beam that
we model as an Airy pattern.  We are interested in the Fourier
transform of the primary beam per unit area, 
\begin{equation}
\label{eqn:beam_uspace}
\tilde{A}(u) = \mathcal{A} \left[\arccos
  \frac{u}{u_\uc} -
  \frac{u}{u_\uc} \sqrt{1-\left(\frac{u}{u_\uc}\right)^2}\,\right],
\end{equation}
where $u \equiv k / (2\pi)$, $\mathcal{A} \equiv 2/(\pi^4 d^2)$, and
$d=6\,\um$.  The beam in Eq.~(\ref{eqn:beam_uspace}) is defined for
$u<u_\uc$, where, at a given wavelength $\lambda$,
$u_\uc=\ang/\lambda$, with $\ang$ the characteristic maximum opening
of the telescope set to $70^\circ$~\cite{Marriage:2006}.  For
$u>u_\uc$, $\tilde{A}(u)=0$.  The normalization is chosen such that,
with the convention
\begin{equation}
A(\vect{r}) = \frac{1}{(2\pi)^2} \int 
  \tilde{A}(\vect{u})\,
  \ue^{2 \mathrm{i} \pi (\vect{u}\cdot \vect{r})}\,
  \ud^2 \vect{u},
\end{equation}
$A(0)=1$.  In the small-angle approximation,
Eq.~(\ref{eqn:beam_uspace}) becomes, in multipole space,
\begin{equation}
\label{eqn:beam_lspace}
\tilde{A}(\ell) = \mathcal{A} \left[\arccos
  \frac{\ell}{\ell_\uc} -
  \frac{\ell}{\ell_\uc} \sqrt{1-\left(\frac{\ell}{\ell_\uc}\right)^2}\,
  \right],
\end{equation}
with $\ell_\mathrm{c} = 2 \pi d/(\lambda \theta)$.  The normalized
primary beam $A(\ell)\equiv \tilde{A}(\ell)/\tilde{A}(0)$ is shown in
the right panel of Fig.~\ref{fig:cls_all} for the three ACT frequency
bands.


\subsection{Generating mixed maps}
\label{sec:mixed}

\begin{figure*}
\begin{center}
\includegraphics[width=7.cm]{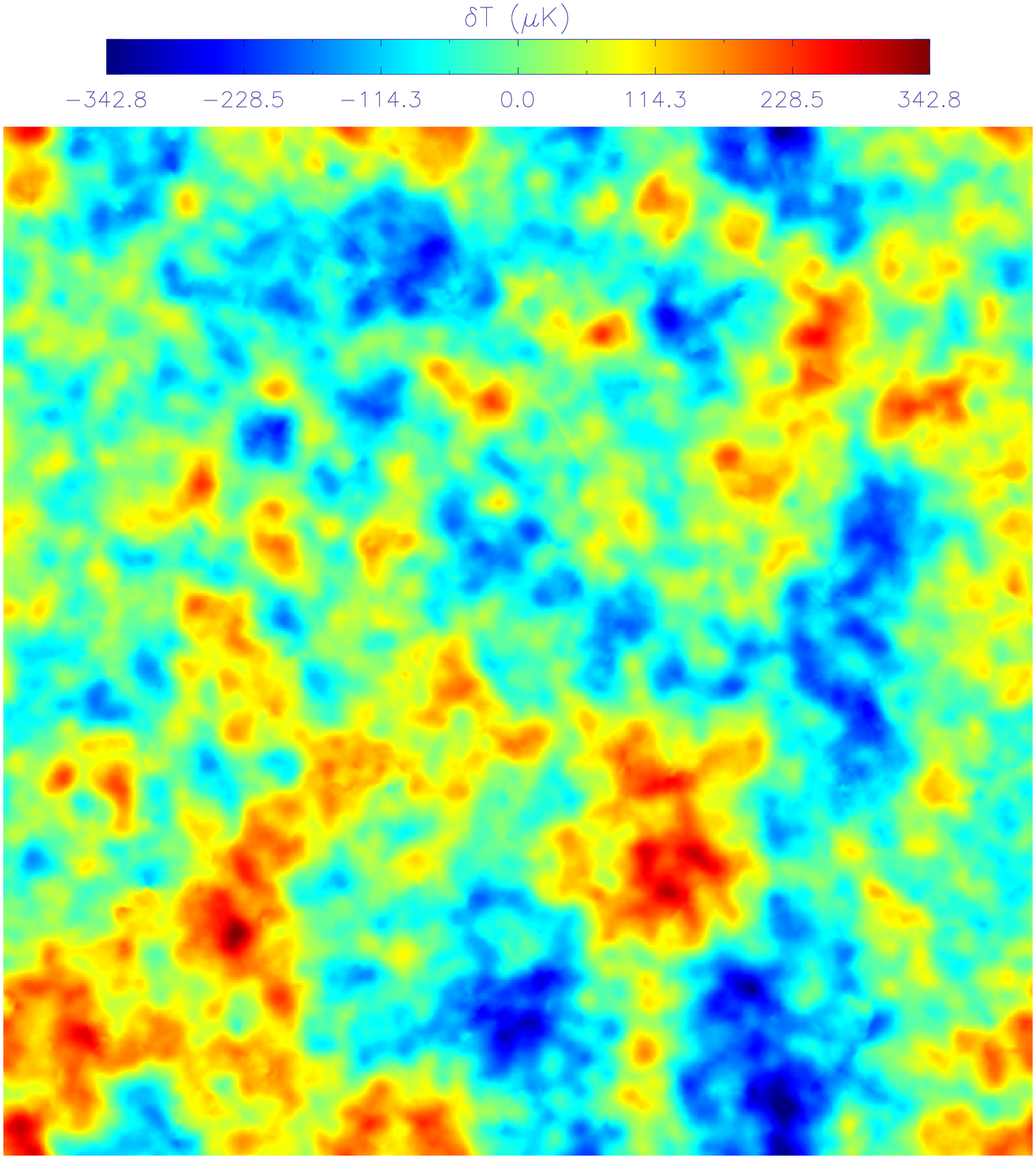}
\includegraphics[width=7.cm]{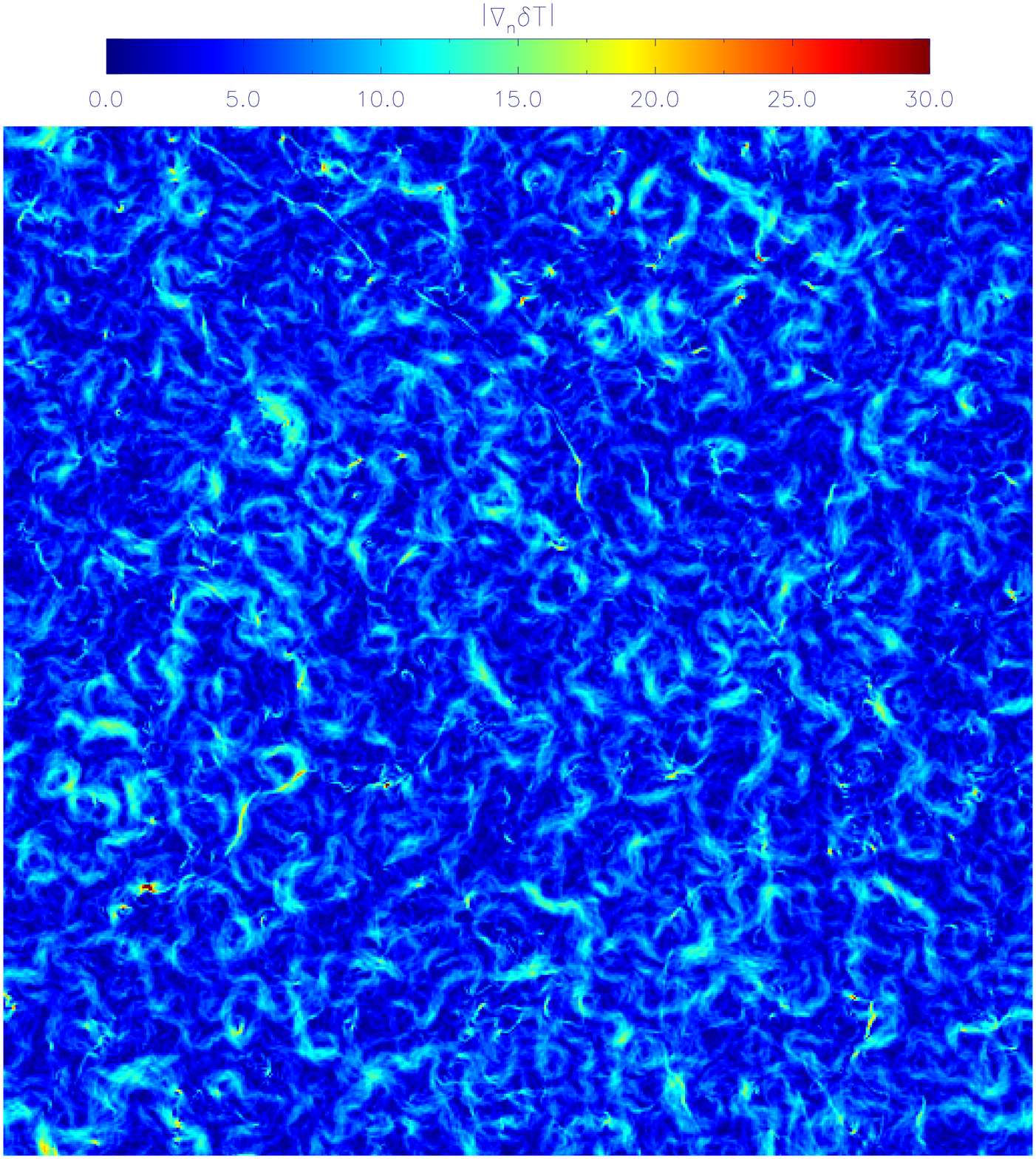}
\caption{CMB temperature anisotropies, including the same primary and
  secondary effects as in Fig.~\ref{fig:mixed_0}, as well as a string
  contribution of $\GU=7\times 10^{-7}$, for a $7.2^\circ$ field of
  view at  $147\,\GHz$ (left), and corresponding normalized
  temperature gradient magnitude map (right).  Although the presence
  of strings can only be guessed in the left panel, very distinctive
  string features appear in the right one.  In addition to sharp
  temperature steps scattered throughout the map, a few bright dipoles
  can be seen in localized string and loop regions with kinks and
  cusps.  Once again, these maps take into account the effect of the
  primary ACT beam as derived in Sec.~\ref{sec:beam}.} 
\label{fig:mixed_d7EM7}
\end{center}
\end{figure*}

In this section, we present small-angle maps of the CMB temperature
anisotropies induced by all of the effects mentioned above.  Producing
these maps turns out to be relatively straightforward for two reasons:
the primary anisotropies of inflationary origin, but also the
secondary ones, namely, the tSZ and kSZ contributions, can be treated
as Gaussian effects, and these fluctuations are uncorrelated with the
string-induced anisotropies.  The first of these two properties allows
us to easily generate maps of the corresponding effects.  Rocha
\emph{et al.} indeed showed in Ref.~\cite{Rocha:2005} that, assuming
statistical isotropy, these maps can be produced by convolving a
normalized Gaussian white noise with the power spectrum of interest.
The fact that we do not need to know anything else than the power
spectrum of these effects should not be a surprise as the two-point
correlation function encodes all the statistical information of
Gaussian effects.  Applying this technique with the spectra shown in
the left panel of Fig.~\ref{fig:cls_all}, we can therefore generate
maps including the effects of all primary and secondary anisotropies.
Now that we have these maps, we can use the second property mentioned
above to include the effect of a network of cosmic strings.  As the
latter is uncorrelated with the primary and other secondary
anisotropies, we indeed only have to add those maps to a map of the
string-induced CMB anisotropies.  The only unknown parameter
controlling the overall amplitude of the string-induced temperature
fluctuations is the string tension, for which we use
$\GU = 7\times 10^{-7}$ (see Sec.~\ref{sec:powspec}).  We are then
left with convolving the resulting map with the primary beam
calculated in Sec.~\ref{sec:beam}.  While the tSZ will produce
non-Gaussian spots, we can ignore them in this initial study.  In
practice, these sources can indeed easily be masked out of CMB maps
when they correspond to bright (typically, $5$-$\sigma$) sources.
Moreover, in the case of multi-frequency experiments, it is possible
to separate thermal SZ effects from thermal CMB distortions.

\begin{figure*}
\begin{center}
\includegraphics[width=7.cm]{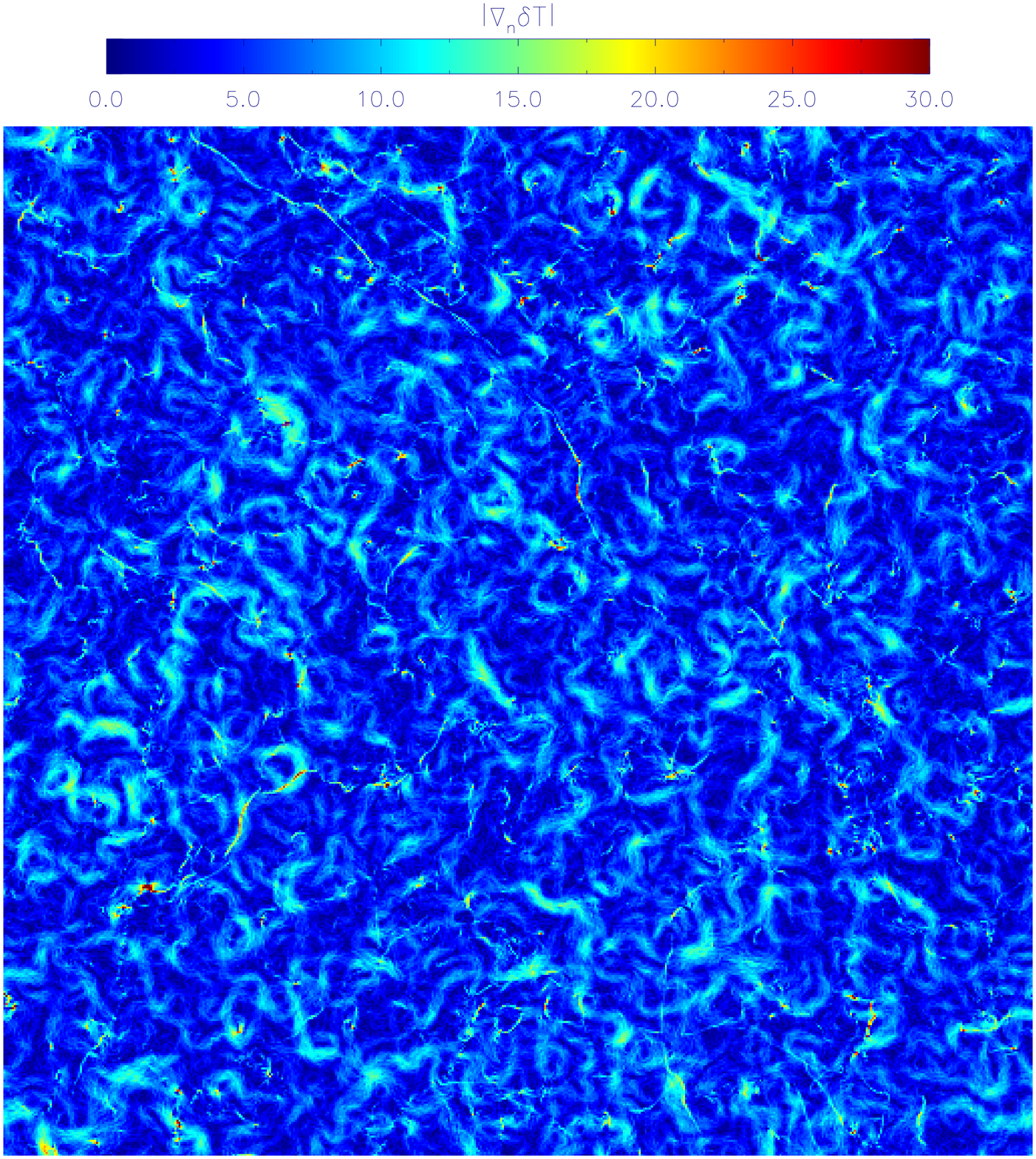}
\includegraphics[width=7.cm]{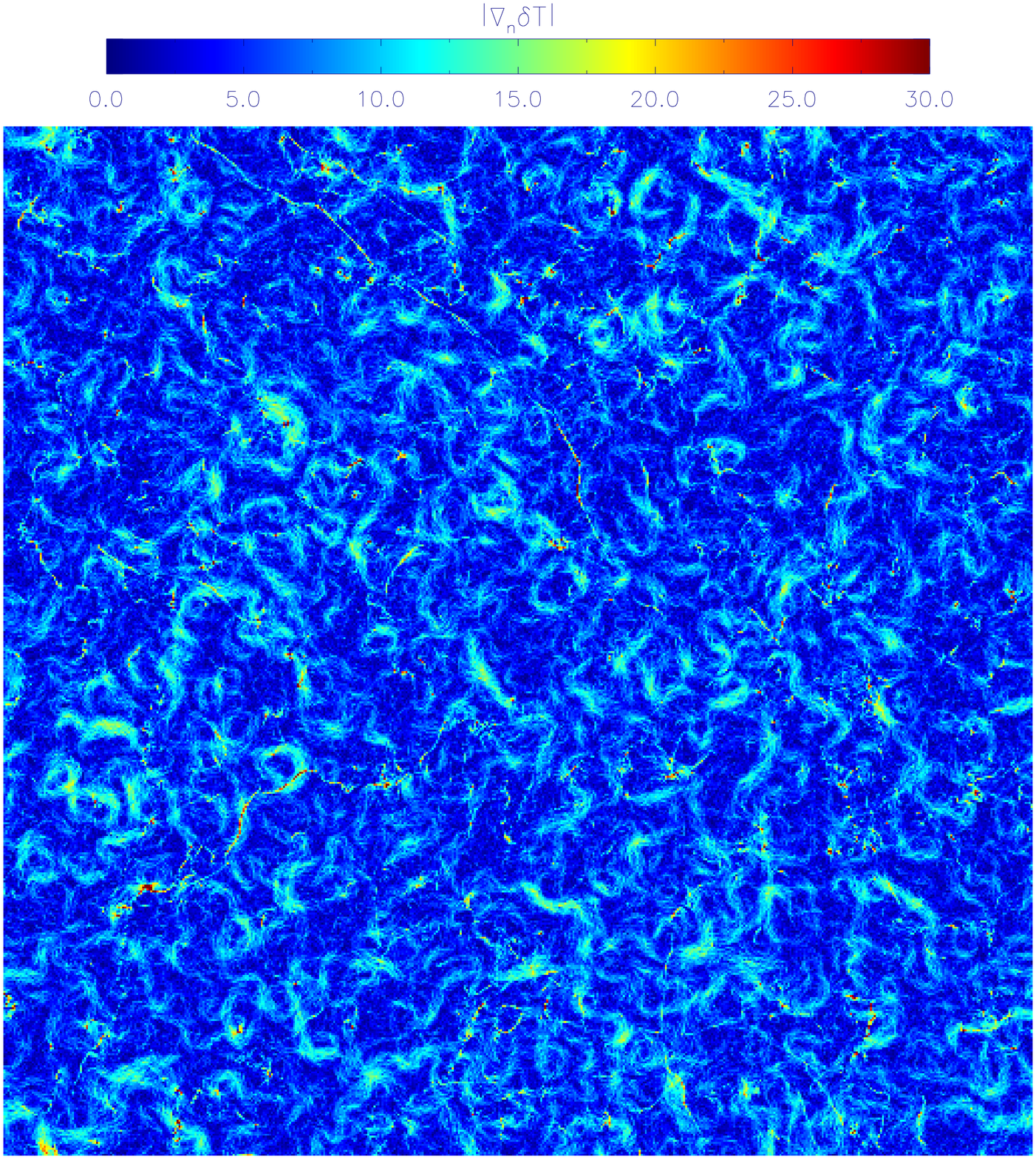}
\caption{Normalized temperature gradient magnitude maps with the same
  components as in Fig.~\ref{fig:mixed_d7EM7}, but at $215\,\GHz$
  (left) and $279\,\GHz$ (right).  The thermal Sunyaev-Zel'dovich
  effect vanishes at $215\,\GHz$ [see Eq.~(\ref{eq:tSZ})].  The
  enhancement of the string signatures with increasing frequency is
  clearly visible when comparing this figure to the right panel of
  Fig.~\ref{fig:mixed_d7EM7}.  Moreover, when the latter is compared
  to the right panel of this figure, filaments produced by the tSZ
  become visible at~$279$~$\GHz$.} 
\label{fig:mixed_ob7EM7}
\end{center}
\end{figure*}

We apply all these techniques to produce maps including the primary
and all secondary anisotropies discussed above, as well as the effect
of a network of cosmic strings.  Figure~\ref{fig:mixed_0} is given as a
reference: it does not include any stringy effect.  Therefore, the
patterns associated with the acoustic peaks are clearly visible, as
well as the Gaussian nature of the fluctuations.  The right panel
shows the corresponding normalized gradient magnitude, as defined in
Eqs.~(\ref{eq:gradmag}) and~(\ref{eq:gradnum}), with $\angres=0.42'$.
By construction, the resulting map is highly non-Gaussian and enhances
the rapid angular variations of the temperature patches of the
fluctuation map.  However, these structures remain relatively smooth
and, unlike the string patterns of Fig.~\ref{fig:gradmag}, do not
exhibit any saturating bright spot.  As a result, we were here able to
use a linear color scale. 

In Fig.~\ref{fig:mixed_d7EM7}, we now include the effect of a network
of strings with $\GU=7\times 10^{-7}$.  The temperature fluctuations
remain dominated by the Gaussian $\Lambda$CDM signal and the secondary
anisotropies, with hints of temperature steps induced by strings.
Note the bright dipole in the bottom left corner of these maps: it is
no longer symmetric in temperature due to the presence of the Gaussian
sources.  Even in the presence of the dominant primordial and
secondary fluctuations, the gradient map still reveals the strings'
location and velocity.  However, the slow-moving segments are now
hidden in the gradient of the Gaussian patterns.  As a result, the
``faint'' strings appear to be discontinuous.  The normalized gradient
color scale, although linear, has been limited to $30$, while some of
the saturating gradient values go up to $46$, as one may have expected
from the PDF shown in the left panel of Fig.~\ref{fig:1pt}.  Finally,
Fig.~\ref{fig:mixed_ob7EM7} shows the gradient magnitude of the
temperature fluctuations generated by the same sources at $215\,\GHz$
(left panel) and $279\,\GHz$ (right panel), whereas the maps shown in
Figs.~\ref{fig:mixed_0} and~\ref{fig:mixed_d7EM7} assume a frequency
of $147$~GHz.  The temperature map is not represented since it is
basically identical to the one of Fig.~\ref{fig:mixed_d7EM7}.  The
gradient color scale is kept the same as in this figure to make the
comparison with the latter easier.  As can be seen on these plots, the
string ``brightness'' increases with increasing frequency, therefore
reflecting the beam frequency dependence shown in the right panel of
Fig.~\ref{fig:cls_all}.  As already mentioned, the increase in angular
resolution accompanying the increase in frequency enhances the
gradient magnitude, but, interestingly, only for the intrinsically
discontinuous anisotropies, such as the ones generated by strings.  At
these frequencies, the highest normalized gradient magnitudes observed
go up to $65$ at $215\,\GHz$ and $95$ for the $279\,\GHz$ channel.  As
can be seen by comparing the right panel of Fig.~\ref{fig:mixed_d7EM7}
to the left panel of Fig.~\ref{fig:mixed_ob7EM7}, increasing the
angular resolution also induces a higher sensitivity to the secondary
sources.  SZ features clearly appear on the $279\,\GHz$ gradient map
as small filaments, whereas they are absent on the $147$~GHz map,
because of the extra smoothing coming from a narrower beam.  The maps
shown in this section therefore suggest that upcoming
arcminute-resolution CMB experiments, such as ACT, could significantly
improve the current constraints on the string tension $GU$.  In a
subsequent paper, we will explore developing optimal non-Gaussian
estimators aimed at detecting the coherent linelike string features
in temperature and gradient maps.  However, they are different enough
from the features produced by the tSZ effect and other known sources
of non-Gaussianity that we can already use a very naive statistic to
get a crude estimate of the detection threshold that the
aforementioned experiments could reach.  Requiring that at least of
order ten linelike features be visible in our gradient maps, we can
clearly see non-Gaussianities down to $GU \simeq 2\times 10^{-7}$.  We
suspect that more detailed studies will yield a firmer detection
limit.


\section{Conclusion}
\label{sec:end}

Based on numerical simulations of Nambu-Goto cosmic string networks,
we produced $84$ high-resolution string-induced CMB temperature maps
over a $7.2^\circ$ field.  We extracted the expected probability
distribution function and the angular power spectrum of string-induced
CMB temperature anisotropies up to $\ell \simeq 10^4$, including
estimates of the systematic and statistical errors associated with our
numerical simulations.  Both of them exhibit distinctive features.  In
particular, strings should induce deviations from Gaussianity due to
rare high temperature fluctuations, and lead to a power spectrum
slowly decaying with increasing multipoles according to
Eq.~(\ref{eq:clsfit}).  Unless $GU$ is well below the inflationary
scale, the string-induced fluctuations start dominating the primary
anisotropies for $\ell>\ell_\mathrm{min}$, with
$\ell_\mathrm{min} \leq 10^4$.  However, for values of $GU$ smaller
than a few times $10^{-7}$, the string-induced anisotropies are
themselves dominated by the tSZ, OV and nonlinear kSZ signals. 

We also discussed the observability of strings in a typical
arcminute-resolution experiment, focusing mainly on maps of the
temperature gradient magnitude.  Including the SZ effects in addition
to the currently favored $\Lambda$CDM temperature anisotropies in our
analysis, we find strings to be ``eye visible,'' in the sense that
at least of order ten string features are observable on a $7.2^\circ$
gradient map, for $\GU$ down to $2\times 10^{-7}$.  As already
mentioned, a drawback in our study is that we do not include the
effect of point sources.  The main risk associated with point sources
could actually be in their removal from CMB maps.  Indeed, for values
of $\GU$ lower than $2 \times 10^{-7}$, the gradient maps no longer
show the typical linelike shape of strings but still contain rare high
temperature fluctuations coming from the active string regions.  Since
the latter are localized, they may look like standard point sources,
which could lead to their exclusion from CMB maps during the
point-source removal process.  A potential disambiguation method might
rely on the intrinsic dipolar signature of these active string regions
in the CMB temperature maps.  Finally, let us mention that being able
to take the gradient of a real temperature map requires a very good
understanding of its noise properties, which could limit the
applicability of the gradient technique described in this paper.

In addition to the statistical analyses we discussed at the end of
Sec.~\ref{sec:mixed}, other extensions of the present work might be
worth thinking of.  In particular, one might want to generate
polarization maps, and especially their B-mode component since they
have been shown to be string tracers~\cite{Seljak:2006, Kunz:2007}.
Some of the results presented here may also be relevant to the
production of gravitational waves by cosmic strings.  The presence of
cusps is indeed expected to trigger gravitational wave bursts.  Since
they also lay on our past light cone, their number is certainly
related to the number of bright dipoles one may observe in our
simulated maps.


\acknowledgments
The authors would like to thank Beth~Reid and Hy~Trac for their help
in computing the tSZ, OV, and kSZ power spectra.  C.R. benefited from
the hospitality of the Department of Astrophysical Sciences at
Princeton University, where this work was originated, and A.A.F. from
a visit to Louvain University, where the final stages of this work
were completed.  The cosmic string simulations have been performed
thanks to computing support provided by the ``Institut du
D\'eveloppement des Ressources en Informatique Scientifique''
(http://www.idris.fr) and the Planck-HFI processing center at the
Institut d'Astrophysique de Paris.  A.A.F. acknowledges support from
Princeton University, NASA Astrophysics Theory Program Grant
No. NNG04GK55G, and NSF Program in International Research and
Education Grant No. OISE-0530095.  This work was also partially
supported by the Belgian Federal Office for Scientific, Technical, and
Cultural Affairs through the Inter-University Attraction Pole Grant
No. P6/11.  We acknowledge the use of the Legacy Archive for Microwave
Background Analysis (LAMBDA).  Support for LAMBDA is provided by the
NASA Office of Space Science.


\bibliography{bibsmap}

\end{document}